\documentclass[twocolumn]{aastex631}

\usepackage{enumerate}[i]
\usepackage{graphicx}
\usepackage{array,multirow,graphicx}
 \usepackage{float}
 \usepackage{supertabular}
 \usepackage{amssymb}
\usepackage{xcolor}

\usepackage[T1]{fontenc}

\DeclareRobustCommand{\VAN}[3]{#2}
\let\VANthebibliography\thebibliography
\def\thebibliography{\DeclareRobustCommand{\VAN}[3]{##3}\VANthebibliography}

\usepackage{longtable}
\usepackage{graphicx}	
\def\code#1{\texttt{#1}}
\usepackage{newtxtext,newtxmath}
\usepackage{rotating}
\usepackage{tikz}
\newcommand{\angstrom}{\mbox{\normalfont\AA}}

\begin{document}

\title[Chemical Doppelgangers in GALAH DR3]{Chemical Doppelgangers in GALAH DR3: the Distinguishing Power of Neutron-Capture Elements Among Milky Way Disk Stars}

\author[0000-0002-0900-6076]{Catherine Manea}
\affiliation{Department of Astronomy, The University of Texas at Austin, 2515 Speedway Boulevard, Austin, TX 78712, USA}

\author[0000-0002-1423-2174]{Keith Hawkins}
\affiliation{Department of Astronomy, The University of Texas at Austin, 2515 Speedway Boulevard, Austin, TX 78712, USA}

\author[0000-0001-5082-6693]{Melissa K. Ness}
\affiliation{Department of Astronomy, Columbia University, Pupin Physics Laboratories, New York, NY 10027, USA}
\affiliation{Center for Computational Astrophysics, Flatiron Institute, 162 Fifth Avenue, New York, NY 10010, USA}

\author[0000-0002-4031-8553]{Sven Buder}
\affiliation{Research School of Astronomy \& Astrophysics, Australian National University, Canberra, ACT 2611, Australia}
\affiliation{ARC Centre of Excellence for All Sky Astrophysics in 3 Dimensions (ASTRO 3D), Australia}

\author[0000-0002-3430-4163]{Sarah L. Martell}
\affiliation{School of Physics, UNSW, Sydney, NSW 2052, Australia}
\affiliation{ARC Centre of Excellence for All Sky Astrophysics in 3 Dimensions (ASTRO 3D), Australia}

\author[0000-0003-1124-8477]{Daniel B. Zucker}
\affiliation{School of Mathematical and Physical Sciences, Macquarie University, Sydney, NSW 2109, Australia}
\affiliation{Astrophysics and Space Technologies Research Centre, Macquarie University, Sydney, NSW 2109, Australia}

\begin{abstract} 
The observed chemical diversity of Milky Way stars places important constraints on Galactic chemical evolution and the mixing processes that operate within the interstellar medium.  Recent works have found that the chemical diversity of disk stars is low.  For example, the APOGEE "chemical doppelganger rate,” or the rate at which random pairs of field stars appear as chemically similar as stars born together, is high, and the chemical distributions of APOGEE stars in some Galactic populations are well-described by two-dimensional models.  However, limited attention has been paid to the heavy elements (Z > 30) in this context.  In this work, we probe the potential for neutron-capture elements to enhance the chemical diversity of stars by determining their effect on the chemical doppelganger rate.  We measure the doppelganger rate in GALAH DR3, with abundances rederived using \textit{The Cannon}, and find that considering the neutron-capture elements decreases the doppelganger rate from $\sim$2.2\% to 0.4\%, nearly a factor of 6, for stars with -0.1 < [Fe/H] < 0.1. While chemical similarity correlates with similarity in age and dynamics, including neutron-capture elements does not appear to select stars that are \textit{more} similar in these characteristics.  Our results highlight that the neutron-capture elements contain information that is distinct from that of the lighter elements and thus add at least one dimension to Milky Way abundance space.  This work illustrates the importance of considering the neutron-capture elements when chemically characterizing stars and motivates ongoing work to improve their atomic data and measurements in spectroscopic surveys.
\end{abstract}


\section{Introduction}\label{sec:intro}
The recent decade has brought forth an exponential increase in available stellar spectroscopic data, enabling population-level analyses of the chemical compositions of Milky Way stars at unprecedented scale.  Massive spectroscopic surveys such as Apache Point Observatory Galactic Evolution Experiment \citep[APOGEE, ][]{apogee}, Large Sky Area Multi-Object Fiber Spectroscopic Telescope \citep[LAMOST, ][]{LAMOST}, Gaia-European Southern Observatory \citep[Gaia-ESO, ][]{Gaia-ESO}, Hectochelle in the Halo at High Resolution \citep[H3, ][]{H3}, the RAdial Velocity Experiment \citep[RAVE, ][]{RAVE}, and GALactic Archaeology with HERMES \citep[GALAH, ][]{GALAHDR3} have provided the Galactic science community with \textit{millions} of stellar spectra.  In the wake of this abundance of stellar spectroscopic data, recent work has begun to investigate \textit{how} much information is actually contained in these datasets.  Does each element carry unique information, or are many of these abundances correlated?  More specifically, do stars at a fixed metallicity tend to display similar chemical profiles in elements across the periodic table, or is there more diversity in their possible compositions?  The literature refers to this notion as the \textit{dimensionality} of chemical abundance space.  

The dimensionality of Milky Way abundance space carries both physical and practical implications that affect a wide range of subfields of astronomy.  When viewed through the physical lens, the dimensionality of Milky Way chemical abundances traces the stability of nucleosynthetic yields across Galactic time and space and the scale and efficiency at which newly synthesized elements are dispersed and mixed into the interstellar medium.  Furthermore, because stellar compositions dictate the architectures and compositions of their planetery systems \citep[e.g.,][]{fischer05, Nielsen2023}, the dimensionality of abundance space dictates the expected diversity of planetary systems around Milky Way stars.  When viewed through the practical lens, the question of chemical dimensionality seeks to assess whether one truly needs to measure tens of elements to fully understand a star's composition, or if measuring a few elements and inferring the rest produces the same quantity and quality of information, thereby significantly enhancing efficiency.  Additionally, the dimensionality of Milky Way abundance space contributes to the validity of strong chemical tagging, the method of reconstructing dispersed stellar birth siblings using chemistry alone \citep[e.g.][]{Freeman2002}.  Strong chemical tagging poses two requirements: (1) that stars born together are chemically homogeneous \citep[e.g.,][]{Bovy2016a, Hawkins2020, Nelson2021} and (2) that groups of stars born together are chemically unique \citep[e.g.,][]{Lambert2016, Price-Jones2020, Cheng2020}.  If chemical dimensionality is sufficiently high, then the strong chemical tagging requirement that each birth cluster possesses a unique chemical profile could be satisfied.  Finally, this topic carries practical implications for extragalactic studies: if chemical abundances are low-dimensional, then one need not spend observing resources sampling entire galaxies if small fields with limited elements sampled are enough to extrapolate the chemical compositions of the remaining stars.

\citet{Ting2012} was among the first works to directly investigate the dimensionality of Milky Way chemical abundance space.  They performed principal component analysis on combined data from several different high-resolution spectroscopic studies to find that stellar abundances tend to possess between six and nine principal components associated with various nucleosynthetic sites, a result they later validate in APOGEE \citep{Ting2022}.  Other works, however, find that chemical abundances may have as few as two dimensions.  For example, \citet{Ness2019} finds that for the low-$\alpha$ disk, with a star's [Fe/H] and age, one can predict its remaining APOGEE abundances to within 0.02 dex.  Furthermore, a star's abundance of elements produced in supernovae can be predicted to $\approx$0.015 dex using Fe, Mg, and age \citep{Ness2022}. This is on the order of the intrinsic scatter of these elements within open clusters, where stars are known to be born together \citep[e.g.][]{Bovy2016a}. This implies that these three dimensions link to birth radii, but individual elements produced in supernovae can not be used to distinguish individual birth groups. \citet{Sharma2022} similarly find that the abundances of most GALAH-reported elements can be inferred to within 0.03 dex using just [Fe/H] and age, though they note that certain elements such as Y and Ba are exceptions to this and cannot be predicted well.  \citet{Weinberg2022} and \citet{ Griffith2022, Griffith2023} also address this question.  They create a two-dimensional (also called a two-zone) model that describes the chemical evolution of the Milky Way according to global enrichment due to time as well as the relative ratio of Type Ia to Type II supernovae.  They then subtract this model from the APOGEE and GALAH data and study the residual abundance patterns in the data.  In APOGEE, the two-dimensional model is suitable enough to predict a star's APOGEE abundances to within 0.03 dex for all well-measured elements \citep{Weinberg2022}, with the addition of a third dimension modelling asymptotic giant branch (AGB) star nucleosynthesis marginally improving the representation of the data \citep{Griffith2023}.  In GALAH, the two-dimensional model produces abundance residuals less than 0.07 dex for most well-measured elements, and the addition of a third dimension, again associated with asymptotic giant branch (AGB) star nucleosynthesis, further decreases abundance residuals  \citep{Griffith2022}.

The majority of the analysis that has been done in the context of APOGEE reports a low-dimensionality of chemical abundance space. However, APOGEE is limited in the nucleosynthetic channels it samples. Therefore, the dimensionality of the Milky Way's chemical abundances is still an open question.  In this work, we seek to address the specific role that the neutron-capture elements play in the chemical dimensionality of Milky Way abundance space.  The GALAH survey offers a more rigorous test of the underlying dimensionality of the Milky Way disk, as it is explicitly designed to capture five channels of nucleosynthetic enrichment for the purpose of testing the validity of strong chemical tagging \citep{DeSilva2015}.  Most critically, due to its optical coverage and high resolution ($R\sim28,000$), GALAH measures the neutron-capture elements.  These include elements formed in both the rapid (\textit{r-}) and slow (\textit{s-}) neutron capture processes, two nucleosynthetic families that are not presently well-measured in APOGEE but may add to the dimensionality of Milky Way abundance space \citep[e.g.][]{Lambert2016, Griffith2022, Sharma2022}.

We study the potentially unique information contained in neutron-capture elements through the lens of "chemical doppelgangers," stars that are highly chemically similar but otherwise dynamically unrelated.  \citet{Ness2018} was the first to measure the so-called "chemical doppelganger rate," defined as the rate at which randomly drawn pairs of field stars are measured to be as chemically similar as stars born together.  Stars that are born together, such as those in open clusters, have been found to be highly chemically similar, with intrinsic dispersions in APOGEE-measured elements ranging from 0.005 to 0.070 dex \citep[e.g.,][]{Bovy2016b, Poovelil2020}.   Thus, open cluster stars are typically considered to represent an upper limit for stellar chemical homogeneity and random field stars the lower limit.  \citet{Ness2018} found that between 0.3 and 1\% of randomly drawn field pairs in APOGEE DR13 \citep{apogeedr13} appear to be as chemically similar as stars residing in open clusters.  While a fraction of these field pairs are likely true conatal pairs from dispersed open clusters, it is unlikely that all are conatal.  To place this result in perspective, if all massive star clusters that have formed in the disk had unique chemical abundance profiles, the expected recovery rate of true birth siblings would be closer to $10^{-4}$ or $10^{-5}$ given the star formation history of the Milky Way \citep[e.g.,][]{BlandHawthorn2010}.  Thus, most doppelgangers are likely not true birth siblings, and stars can share remarkably similar chemical profiles despite being born of different star forming complexes due to a relatively homogeneous chemical evolution of the thin disk. These results are validated by \citet{deMijolla2021} in APOGEE DR16 \citep{apogeedr16}.  Together, these works suggest that the diversity of Milky Way disk star abundances is qualitatively low.

In this work, we perform the first measurement of the doppelganger rate in GALAH.  We center our investigation on whether the neutron-capture elements affect the measured doppelganger rate.  If neutron-capture elements add at least one unique dimension to chemical abundance space at the available abundance precision, then we expect the doppelganger rate to decrease with the addition of this nucleosynthetic family.  If neutron-capture elements do not add a unique dimension, then we expect the doppelganger rate to remain relatively unchanged with the addition of these elements.  Through this test, we implicitly probe whether neutron-capture elements enhance the diversity of Milky Way stars or simply trace the lighter (light/odd-Z, $\alpha$, and iron-peak) elements primarily produced in supernovae.  We use the following questions to guide our analysis:

    \begin{enumerate}[i]
        \item What is the doppelganger rate in GALAH?
        \item How does the inclusion of neutron-capture elements specifically affect the doppelganger rate?
        \item Do there exist pairs of stars that are "doppelganger" in the light, $\alpha$, and iron-peak elements but show differences in the neutron-capture elements?  If so, are there any physical characteristics that differentiate these pairs from pairs that are "doppelganger" in all elements?
    \end{enumerate}

In Section \ref{sec:data}, we describe the GALAH dataset and our choice of open cluster stars that serve as a reference in our doppelganger rate measurements.  In Section \ref{sec:method}, we use \textit{The Cannon} to re-derive abundance ratios in 16 elements (Fe, O, Al, Mg, Ca, Si, Cr, Cu, Zn, Mn, Zr, Y, Ba, Ce, Nd, and Eu) for the purpose of enhancing precision and ensuring well-constrained uncertainties.  We then measure the doppelganger rate using two distinct approaches.  In Section \ref{sec:results}, we investigate the impact of the neutron-capture elements on the measured doppelganger rate and compare the physical characteristics of stars that are partial doppelgangers (doppelganger in the light,$\alpha$, and iron-peak elements) with those that are complete doppelgangers (doppelganger in all measured elements).  We discuss our results in the context of previous observational and simulation work in Section \ref{sec:discussion} and conclude in Section \ref{sec:conclusion}.

\section{Data}\label{sec:data}
GALAH Data Release 3 \citep[DR3][]{GALAHDR3} serves as the basis of our investigation.  GALAH is an optical (4710 $\rm \angstrom < \lambda$ < 7890 $\rm \angstrom$ spread across four non-contiguous cameras), magnitude limited (V\textless14) spectroscopic survey with high resolving power ($R \sim 28, 000$) that targets Milky Way stars at |b| \textgreater~10$^{\circ}$ \citep{DeSilva2015}.  GALAH DR3 reports stellar parameters (e.g., $\rm T_{eff}$, log g, spectral broadening, [Fe/H], etc.) and abundances for nearly 600,000 stars derived using Spectroscopy Made Easy (SME, \citealp{Valenti1996, Piskunov2017}) and 1D MARCS model atmospheres \citep{Gustafsson1975, Bell1976, Gustafsson2008}.  Non-local thermodynamic equilibrium (NLTE) is assumed during spectral line synthesis of 13 elements (H, Li, C, O, Na, Mg, Al, Si, K, Ca, Mn, Fe, and Ba) whereas LTE is assumed for the rest.  For each star, GALAH DR3 reports its surface abundances in up to 30 elements spanning five major nucleosynthetic channels: the light (Li, C), $\alpha$ (Mg, Si, Ca, O), odd-Z (Na, Al, K), iron-peak (Sc, V, Cr, Mn, Fe, Co, Ni, Cu, Zn, Ti), and slow (\textit{s}-) and rapid (\textit{r}-) process neutron-capture (Rb, Sr, Y, Zr, Mo, Ru, Ba, La, Ce, Nd, Sm, and Eu) elements.  In addition to the main catalog, DR3 provides the community with several other data products, such as a 1-dimensional, radial-velocity-corrected, continuum-normalized, combined spectrum for nearly every star sampled by the survey as well as a catalog of dynamical parameters and age estimates for nearly all GALAH targets.  Dynamical parameters are determined using Python package \code{galpy} (\citealp{Bovy2015}, see GALAH survey webpage\footnote{\code{https://www.galah-survey.org/dr3/the\_catalogs/}} for details on their assumed Galactic potentials and properties).  Age estimates for GALAH stars are determined using Bayesian Stellar Parameter Estimation code (BSTEP) from \citet{Sharma2018}, which uses PARSEC release v1.2S + COLIBRI stellar isochrones \citep{Marigo2017} and Bayesian estimation to infer intrinsic stellar parameters from observables $\rm T_{eff}$, log g, [Fe/H], [$\alpha$/Fe], parallax, and 2MASS J \& H-band magnitudes.

For this investigation, we aim to measure the doppelganger rate using abundances that we re-derive from the GALAH DR3 spectra using \textit{The Cannon} \citep[][see Section \ref{subsec:whycannon} for our motivation for this choice]{Cannon}.  We elect to perform our investigation on red giant and red clump stars in the GALAH survey, as they are on average brighter and thus probe a larger Galactic volume relative to dwarfs.  This enables us to a) better compare to \citet{Ness2018}, which also used giants, and b) understand the doppelganger rate on a broader spatial scale, expanding on work such as that of \citet{Bedell2018} which investigated the chemical diversity of stars within the local (100 pc) Solar neighborhood.  To build our stellar sample, we apply a series of selections that we motivate in the following paragraph:

\begin{enumerate}[i]
	\item \code{flag\_sp} = 0
	\item \code{flag\_fe\_h} = 0
    \item \code{snr\_c2\_iraf} > 20
    \item \code{ruwe} < 1.2
    \item 1.5 < \code{log g} < 3.5
    \item 0.0033*\code{teff} - 13.6 < \code{log g} < 0.0036*\code{teff} - 13.9
    \item -1.20 < \code{fe\_h} < 0.20
    \item -0.25 < \code{Cr\_fe} < 0.15
    \item \code{Cu\_fe} > -0.30
    \item \code{Zr\_fe} < 0.60
    \item \code{Y\_fe} < 0.60
    \item \code{Ba\_fe} < 0.80
    \item -0.50 < \code{Ce\_fe} < 0.40
    \item \code{Nd\_fe} < 0.60
    \item \code{Eu\_fe} < 0.60
\end{enumerate}

Though we re-derive abundances using \textit{The Cannon}, the first two requirements ensure reliable GALAH-reported stellar parameters and [Fe/H] abundances which consequently eliminates clearly problematic spectral data.  The third requirement ensures that all spectra have a signal-to-noise (SNR) ratio above 20 in the $\sim$ 5700 Angstrom spectral region, and the fourth requirement filters for potential spectroscopic binaries \citep[e.g.,][]{ruwe} that were missed by the \code{flag$\_$sp} flag. Requirements (v) and (vi) ensure we select red giant and red clump stars.  The remaining requirements excise certain stars that have either extreme abundances (e.g., \textit{s}-process enhanced stars, which are likely post-mass transfer systems and no longer reflect their natal composition) and/or GALAH-reported abundances that are not sampled by our high-quality training set and thus less reliably inferred by our model (see Section \ref{subsubsec:trainingset}).

\subsection{Open Cluster Catalog}\label{subsec:OC_CATALOG}
As mentioned in Section \ref{sec:intro}, we define doppelgangers to be pairs of field stars that appear as chemically similar to one another as stars residing in open clusters.  As such, a reliable set of reference open cluster stars is critical for our investigation.  We use the open cluster catalog of \citet[][]{Spina2021} to build this reference set.  \citet[][]{Spina2021} builds off of the widely-used open cluster catalog of \citet[][]{Cantat-Gaudin2018}, specifically improving cluster membership determinations for GALAH-sampled open clusters.  They make use of a Support Vector Machine classifier (see their footnote 2 for a detailed description) to re-assess cluster memberships using Gaia astrometry \citep{GaiaEDR3} and validate their results with careful inspection of the resulting cluster isochrones and radial velocity distributions.  For each cluster star, they report a membership probability.   We only consider open cluster stars that have a probability of membership that exceeds 50\% ($\rm P_{mem} > 0.5$). In practice, the vast majority of open cluster stars in our selection have membership probabilities between 90\% and 100\%, but this selection allows for a few additional stars with membership probabilities of 75\%.

\section{Method}\label{sec:method}
The primary goal of this work is to measure how often random pairs of field stars sampled by GALAH appear as chemically similar as GALAH stars in open clusters, referred to as the doppelganger rate.  Throughout this work, we closely follow the method of \citet{Ness2018}.  They compute the doppelganger rate using 20 elements ( Fe, C, N, O, Na, Mg, Al, Si, S, K, Ca, Ti, V, Mn, Ni, P, Cr, Co, Cu, and Rb) homogeneously derived from APOGEE DR13 spectra using \textit{The Cannon} \citep{Ness2015}.  To compute the doppelganger rate, they draw random pairs of field stars unassociated with known clusters and compare their chemical similarity to that of random stellar pairs drawn from within open clusters, which they refer to as \textit{intracluster pairs}.  To quantify the abundance similarity of stars in a pair, they compute a $\chi^2$ value for each pair, defined as:
\begin{equation}\label{eq:chisq}
\chi^2_{nn'} = \sum_{i=1}^I \frac{[x_{ni} - x_{n'i}]^2}{\sigma^2_{ni} + \sigma^2_{n'i}}
\end{equation}
where the two stars in the pair are indexed as $n$, $n'$ and $x$, $\sigma$ are their derived abundance and abundance uncertainty in element $i$.  This leads to a global chemical similarity metric for each pair that considers all sampled elements.  Doppelgangers are defined as stellar pairs with $\chi^2$ values less than the median of intracluster pairs.

\subsection{The Cannon}\label{subsec:Cannon}
As in \citet{Ness2018}, we measure the doppelganger rate using abundances and abundance uncertainties re-derived using \textit{The Cannon} \citep{Cannon}.  \textit{The Cannon} is a data-driven method for determining parameters and abundances from stellar spectra. \textit{The Cannon} does not explicitly use atomic physics to determine these parameters: instead, it fits a suitably flexible model to the relationship between each spectral pixel's intensity and each input label (e.g., $\rm T_{eff}$, log g, [Fe/H], etc.) using a high-fidelity \textit{training} set. We use a second-order polynomial to model the relationship between spectral pixels and the following labels: $\rm T_{eff}$, log g, broadening velocity,  O, Mg, Al, Si, Ca, Cr, Cu, Zn, Mn, Zr, Y, Ba, Ce, Nd, and Eu.  We subsequently infer these labels at \textit{test} time in our implementation. This model is similarly used in several other works that require high precision abundances for hundreds of thousands of stars \citep[e.g.,][Walsen et al. 2023, submitted]{Buder2018, Wheeler2020}.  \textit{The Cannon} is a generative model, and constructs, for each label inference, a probability distribution function for the observed flux - that is, a theoretical spectrum for each star for which the labels are inferred. This enables the goodness-of-fit to be evaluated for the model spectrum versus the data, for each label and for each star. We direct the reader to \citet{Ness2015} for a thorough description of the methodology of \textit{The Cannon}.

\subsubsection{Why Re-Derive GALAH DR3 Abundances Using \textit{The Cannon}?}\label{subsec:whycannon}

At the core of our investigation is the determination of the chemical similarity of pairs of stars.  In an ideal setting, we could simply take the absolute difference in the elemental abundance ratios of each star in a pair to determine their degree of chemical similarity.  However, this is impossible because all abundance measurements have an associated uncertainty, and thus, we must factor them into our determinations of chemical similarity.  In this work, \textit{The Cannon} is critical for 1) providing improved precision of the measured abundances and 2) providing accurate uncertainty estimates for stars that we can validate are well fit by our model in spectral space.  

Abundance precision is a key extrinsic factor that influences the measured doppelganger rate.  With large uncertainties in [X/Fe], it is more difficult to distinguish true doppelgangers from stars that show chemical differences unresolved at the current precision level.  As such, maximizing precision is key to accurately constraining the doppelganger rate.  As explained in \citet{GALAHDR3}, \textit{The Cannon} is capable of outperforming the GALAH DR3's SME-derived abundance precision.  For GALAH DR3, elemental abundances were derived using on-the-fly spectral synthesis, and in order to limit computation time, the syntheses were only performed for a selection of unblended spectral regions associated with each element of interest.  This means that some elemental abundances, such as [Mg/Fe], were derived using just one or two spectral lines, leading to larger uncertainties \citep[e.g.,][]{Jofre2019}.  \textit{The Cannon} in part achieves increased abundance precision  because it leverages the \textit{entire} spectrum to retrieve abundance information. This means it can use \textit{all} spectral features -- strong lines, weak lines, blended lines, lines with uncertain line data, and even continuum effects -- associated with each element to infer an abundance.  It is known that changes in the chemical composition of a star's atmosphere will affect the atmospheric opacity profile of the star, particularly when the abundance of electron-donating atoms is altered \citep{Ting2018}.  Changes in opacity will consequently affect several different parts of the spectrum, including the continuum and line strengths of other elements.  \textit{The Cannon} is able to capture these trends, though careful consideration of non-physical correlations between abundances and spectral features must also be taken (see Section \ref{subsubsec:trainingset}).

Well-constrained abundance uncertainties are also key in this investigation. Underestimated uncertainties will artificially decrease the doppelganger rate while over-estimated uncertainties will artificially increase it.  Abundance uncertainties from classical abundance determination methods are influenced by a collection of sources, including but not limited to uncertainties in the data reduction process, input atomic physics, and choice of continuum placement when fitting a spectrum \citep[e.g.,][]{Jofre2017}.  Using \textit{The Cannon}, we can determine a systematic cross-validation uncertainty for each label, that represents the fidelity with which we recover the training labels. This measurement uncertainty incorporates the uncertainty on the training labels that are inherited from GALAH. However, we can take advantage of repeat visits of individual stars to quantify our internal precision, which represents the overall systematic precision with which we can determine each stellar label, and this is what we ultimately adopt for our abundance uncertainties (see Section \ref{subsubsec: uncerts}). It is important to highlight that in our investigation, abundance accuracy, which we inherit from the training set of stars and can not control, is not important.  Our only requirement is that spectra that are identical possess identical labels, so any global offsets in abundance do not impact our result, only relative offsets between different stars due to differing chemistry.

\subsubsection{Re-derived Stellar Parameters and Abundances}\label{subsec:abundsofinterest}
Instead of re-deriving the full array of elements reported by GALAH, we select a subsample of the elements. We select elements that enable us to ask our scientific question of the data but limit our selection to the set of abundances for which we can build a high-fidelity training set.  Our selection samples elements from each major nucleosynthetic family, including  a light/odd-Z element (Al), $\alpha$ elements (O, Mg, Ca, Si), iron-peak elements (Fe, Cr, Cu, Zn, Mn), first (\textit{s}-process) peak elements (Y, Zr), second \textit{s}-process peak elements (Ba, Ce, Nd), and an \textit{r}-process element (Eu).  In addition to these elements, we re-derive $\rm T_{eff}$, log g, and \code{v\_broad} (broadening velocity due to rotation, macroturbulence, etc.) for each spectrum.

\subsubsection{Additional Modifications to Spectra Prior to Input into \textit{The Cannon}}
Upon downloading\footnote{GALAH DR3 spectra can be downloaded from \code{https://datacentral.org.au/services/download/}} all GALAH DR3 spectra that satisfy the conditions enumerated in Section \ref{sec:data}, we interpolate each star's flux and flux error array over the shared wavelength grid recommended by the GALAH team.\footnote{\code{https://github.com/svenbuder/GALAH\_DR3/}}  We perform several tests to assess how manipulating the spectra affects our resulting label precision and find that truncating the spectra to only include the first three CCDs ($\lambda \leq$ 6730) increases the performance of \textit{The Cannon}.  This increase in performance from neglecting the last CCD is likely due to the strong spikes in the redmost spectral segment that do not originate in the stellar photosphere and make fitting a model to that spectral region difficult.  These spectral spikes in the redmost CCD are likely due to imperfect telluric subtraction, a common challenge in the spectral reduction of near-infrared spectra \citep[e.g.,][]{Griffith2022}.  We are able to neglect this final CCD without a loss of precision because most of the spectral lines associated with our sampled elements lie in the first three CCDs.  The only exception to this is O: by removing the fourth CCD, we lose access to the only available O lines in GALAH.  Previous works using \textit{The Cannon} have successfully recovered O abundances without using O lines \citep{Ting2018}, so we proceed with inferring O, but we \textit{do not} consider it in our subsequent measurement of the doppelganger rate.  Additionally, we remove three spectral segments containing strong diffuse interstellar bands (DIBs, e.g., \citealp{Vogrin2023}) near $\lambda$5798, $\lambda$5871, and $\lambda$6614.  These features are caused by interstellar dust absorbing the star's light and are not intrinsic to the stellar photosphere.  Thus, to ensure that \textit{The Cannon} does not use these features for inference, we remove them.

\subsubsection{Training Set}\label{subsubsec:trainingset}

\begin{figure}
    \centering
    \includegraphics[width=8cm]{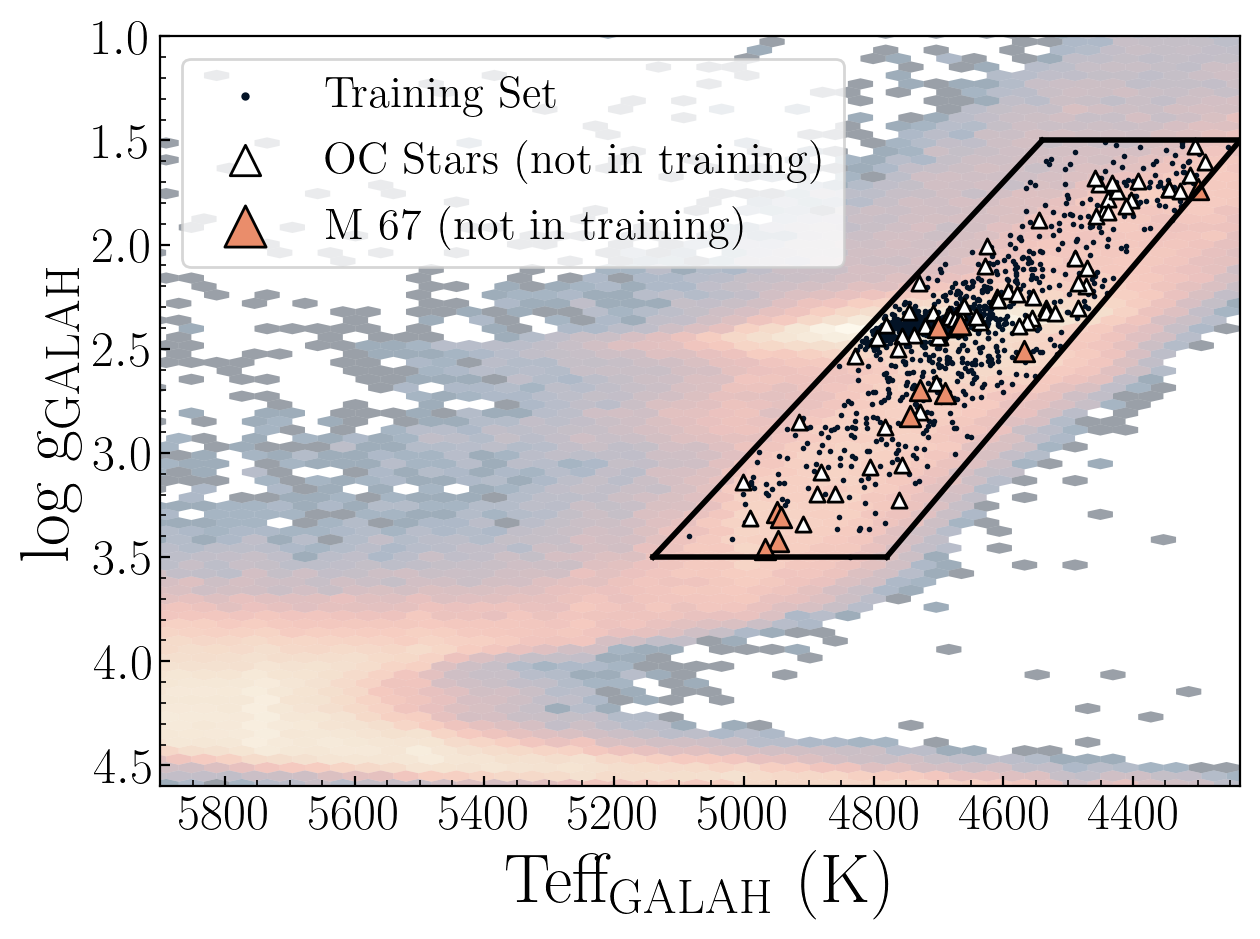}
    \caption{Kiel diagram (GALAH-reported $\rm log g~vs.~T_{eff}$) for the full GALAH sample (background distribution).  Our parameter space of investigation, which was designed to contain red giant and red clump stars, is encapsulated within the black polygon.  Black dots mark stars in our training set.  We include open cluster stars as triangles to illustrate their parameter space coverage, with filled orange triangles highlighting members of chemically-homogeneous open cluster M 67.}  All background stars that fall within the black polygon and satisfy our quality cuts (Section \ref{sec:data}) are considered in our analysis.
    \label{fig:training_set}
\end{figure}

\begin{figure}
    \centering
    \includegraphics[width=8cm]{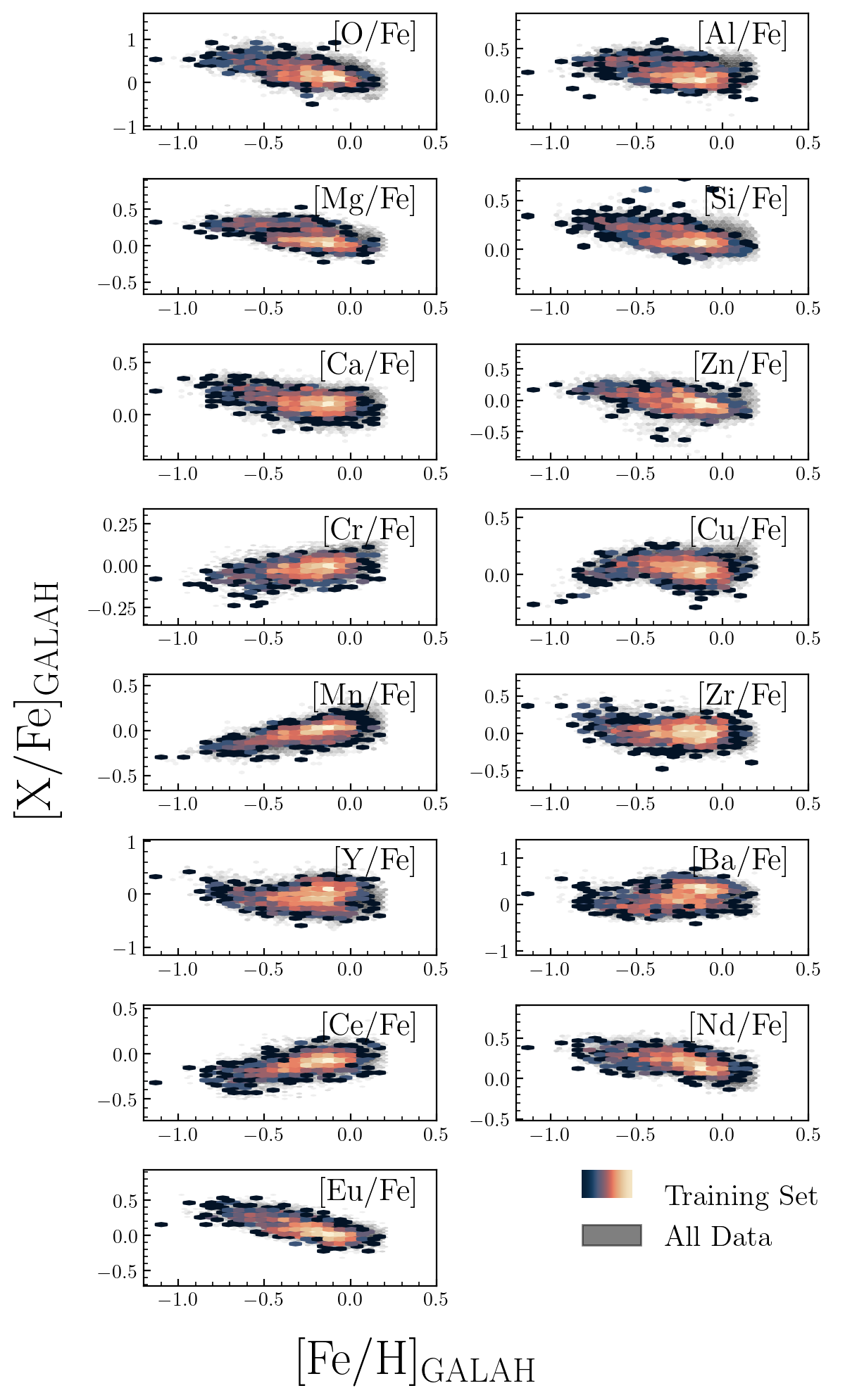}
    \caption{Density distributions in [X/Fe] vs. [Fe/H] for our training sample (color) atop our full sample (background gray).  Note that the colormaps are colored by logarithmic stellar density.  We ensure that the training set spans the parameter space of our full sample to ensure reliable output \textit{Cannon} abundances.}
    \label{fig:x_fe_h_train}
\end{figure}

When building our high-fidelity training set, we experiment with various selections to identify one that allows for high output abundance precision while also sampling sufficient open cluster stars.  Our final choice of training set is limited to stars that satisfy the following requirements, in addition to those mentioned in \ref{sec:data}:
\begin{enumerate}[i]
	\item \code{flag\_X\_fe} = 0
    \item \code{chi2\_sp} < 10
    \item \code{snr\_c2\_iraf} > 100
\end{enumerate}
The first requirement makes use of an element-specific flag that controls for unreliable GALAH-measured abundances in specific elements.  The second requirement reports the $\chi^2$ fit of the best-fitting SME model to the GALAH data.  Previous works, such as \citet{Nandakumar2022}, have required that \code{chi2\_sp} < 4 when building a GALAH-based training set.  We found that increasing our requirement to \code{chi2\_sp} < 10 enables us to achieve the same precision while allowing better sampling of high-metallicity ([Fe/H] > 0) red giant stars.  The final requirement ensures that we are training \textit{The Cannon} on high SNR data to maximize its ability to learn real correlations between spectral pixels and labels as opposed to learning from noise.

In Figure \ref{fig:training_set}, we present a Kiel (log g vs. $\rm T_{eff}$) diagram of our final training set (black dots) consisting of 956 stars that sample the surface area of our chosen parameter space, which we define by the polygon in Figure \ref{fig:training_set}. We note that our chosen polygon is not parallel to the red giant branch.  This means that we are unable to sample cooler, metal-rich giants beyond [Fe/H] = 0.2.  This is out of necessity: there are not enough high SNR stars with unflagged abundances to extend our training sample to this region of the Kiel diagram.  However, this is not a problem as we conduct our entire analysis in the chosen polygon, and the polygon is well-sampled by the training set.  To illustrate the parameter space sampled by our open cluster stars, we include them in the figure as open triangles, with M 67, a chemically homogeneous open cluster \citep[e.g.,][]{Bovy2016a, Ness2018, Poovelil2020}, as orange filled triangles.  We note that these open cluster stars are not in our training set.  The background distribution represents the full GALAH dataset, and all background stars that fall within the polygon and satisfy our quality cuts are re-analyzed using \textit{The Cannon}.  In Figure \ref{fig:x_fe_h_train}, we plot [X/Fe] vs. [Fe/H] distributions for our training set stars atop the equivalent for our full sample, highlighting that the surface area of the [X/Fe] vs. [Fe/H] distribution of our full sample is fully covered by our training set.  This is important for ensuring that the model need not extrapolate when inferring stellar abundances.

\begin{figure*}
    \centering
    \includegraphics[width=\textwidth]{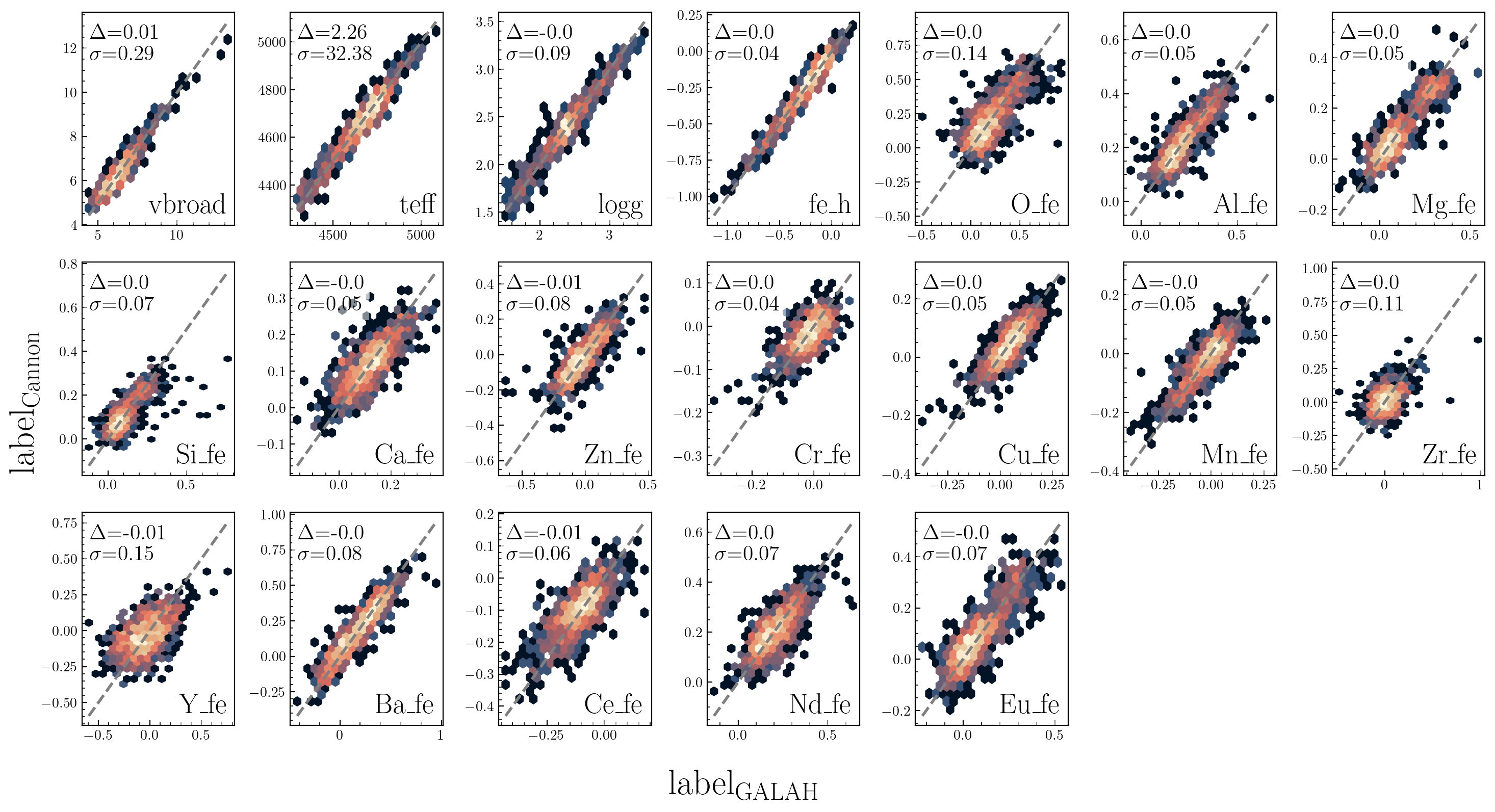}
    \caption{ Combined results of our 10 leave-10\%-out cross-validation tests, where we plot output \textit{Cannon} label as a function of input GALAH DR3 label for all 19 labels-of-interest in our 956 star training set. Colormap corresponds to point density and dashed-line represents the one-to-one line.  The offset from ($\Delta$) and scatter ($\sigma$) around the one-to-one line is printed in the corner of each panel.  Our model is able to retrieve the GALAH-reported abundances of our training set within 0.07 dex for the majority of elements.}
    \label{fig:cannon_performance}
\end{figure*}

\begin{figure*}
    \centering
    \includegraphics[width=\textwidth]{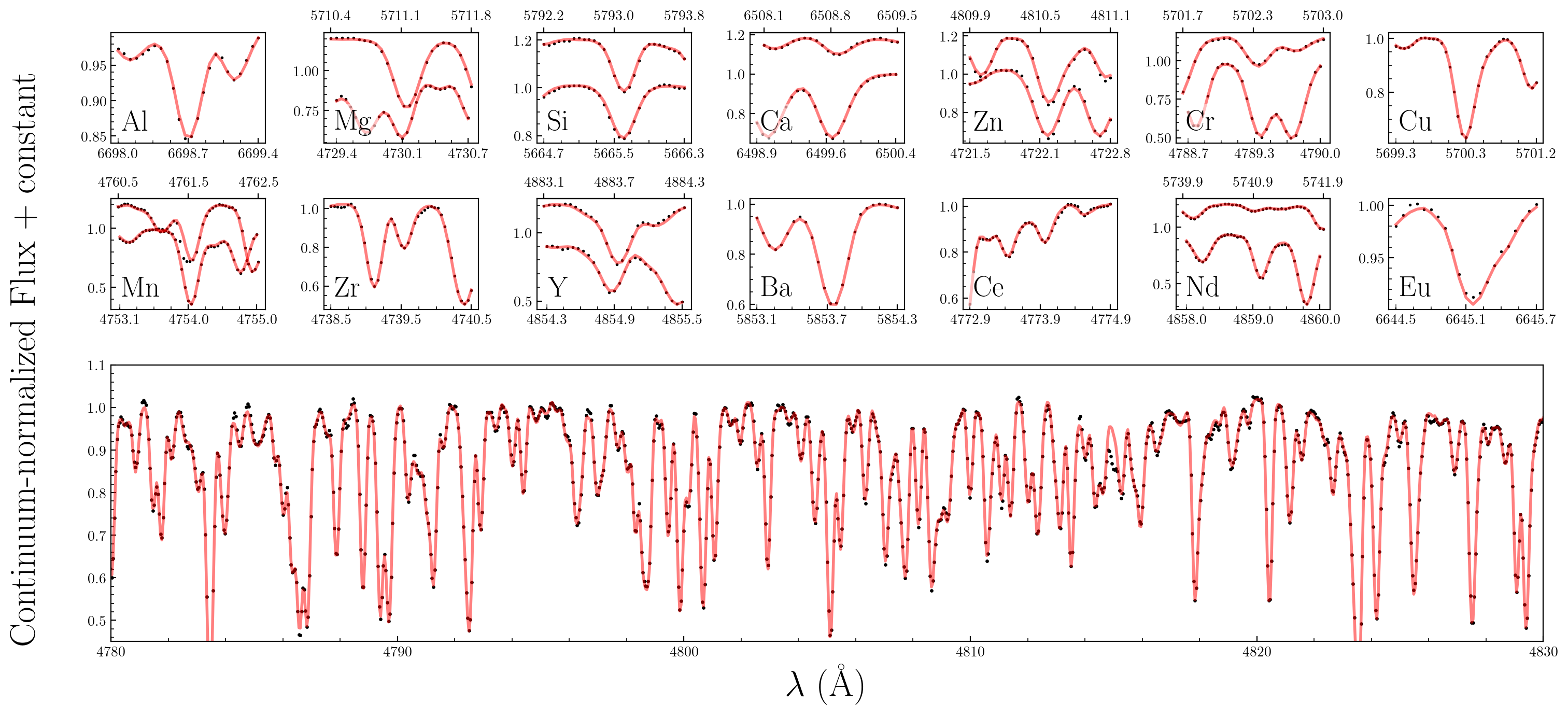}
    \caption{A comparison of our \textit{Cannon} model fit (red) to the spectrum of a star in M 67 (black).  The smaller top panels highlight segments of the spectra corresponding to strong lines of each element.  Bottom panel shows a larger cutout of the spectrum.  It is apparent that \textit{The Cannon} is capable of fitting the GALAH data well, both globally and around lines of interest.}
    \label{fig:cannon_modelvsspec}
\end{figure*}

We assess the ability for our model to recover the GALAH labels of the training set by performing a series of ten leave-10\%-out cross-validation tests.  This involves training our model on 90\% of the training data and assessing its ability to recover the GALAH-reported labels of the remaining 10\% of the training data.  In Figure \ref{fig:cannon_performance}, we plot the \textit{Cannon}-recovered label as a function of input GALAH label for all stars in our training set, marking the one-to-one line for reference.  The model is successful in recovering the input training data labels to high precision, with recovered labels agreeing with GALAH-reported labels within 0.04 to 0.08 dex for most elements. Exceptions to this are O, Zr, and Y, which we recover to within 0.11 to 0.14 dex.  We note that this cross validation is an assessment of the fidelity with which we can determine the reference labels, but as it includes the GALAH label uncertainties, it is not a measurement of the internal precision of \textit{The Cannon} on these data (see Section \ref{subsubsec: uncerts}).  As a measure of robustness, we assess the fit of the output \textit{Cannon} model spectra to the input GALAH spectrum, both globally and around strong lines of measured elements, via a $\chi^2$ goodness-of-fit metric that considers GALAH flux uncertainties.  To determine goodness-of-fit to specific lines, we adopt the GALAH SME line masks presented in \citet{Buder2018}.  We flag and subsequently ignore all stars with global $\chi^2$ and line-specific $\chi^2$ values that exceed two times the degrees of freedom (e.g., spectral pixels).  In Figure \ref{fig:cannon_modelvsspec}, we show an example fit to a typical open cluster star to illustrate the quality of \textit{The Cannon}'s model spectra fits.

\begin{figure*}
    \centering
    \includegraphics[width=17cm]{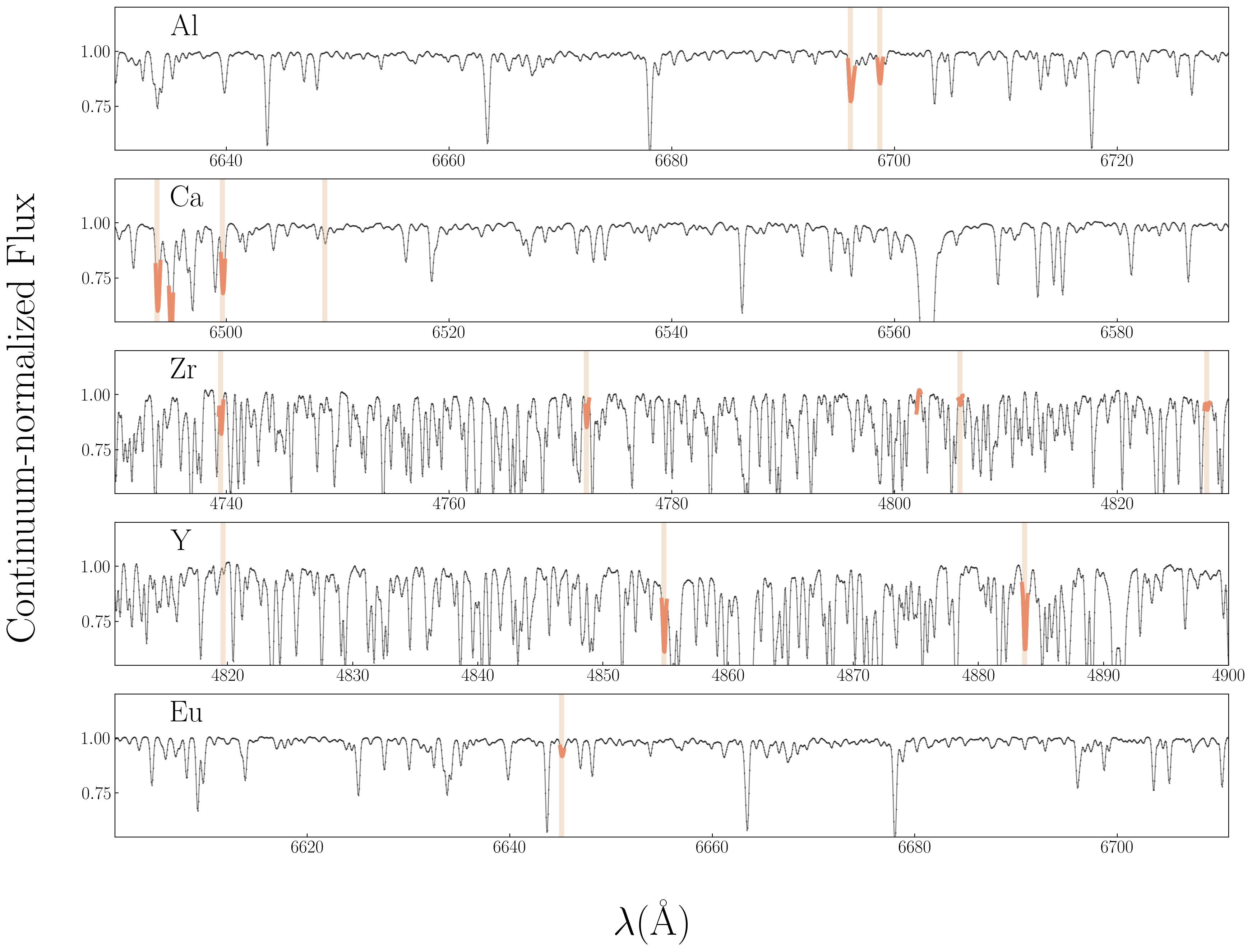}
    \caption{Spectral windows, showing the continuum normalized flux as a function of wavelength, for the median stellar spectrum in our sample (black) containing regions of absorption lines of Al, Ca, Zr, Y, and Eu, from top to bottom.  We highlight in orange spectral regions that correspond with the 1\% largest first order \textit{Cannon} model coefficients in the labelled element.  The black spectrum is the base \textit{Cannon} spectrum, which represents the median spectrum of the full dataset.  Light orange vertical lines correspond to known lines of the element, and thin black vertical lines correspond to known lines of Fe.  We note that, even in the most difficult to measure elements such as Zr, the \textit{Cannon} model successfully draws its primary abundance information from the relevant element's lines. See Appendix Figure A\ref{fig:coeffs} for the full array of first order coefficients as a function of wavelength for each label.}
    \label{fig:speccan}
\end{figure*}

When using data-driven algorithms such as \textit{The Cannon} to measure abundances from the full spectral range without the use of censoring, a procedure where \textit{The Cannon} is only allowed to learn abundance information from specified strong lines of each element, it is important to acknowledge that the model can infer abundances using correlations between spectral features not directly associated with the element. As mentioned in Section \ref{subsec:whycannon}, in some cases, there are physical reasons for the existence of these correlations, as changes in a star's atmospheric composition in an element can influence the spectral behavior of other elemental lines or continuum regions \citep[e.g.,][]{Ting2018}.  However, in other cases, this can lead to abundance inferences of certain elements that are instead primarily driven by a non-physical correlation that is introduced by the training set data.  In the context of our high resolution spectra, we expect the primary abundance information for each element to be learned from strong, known lines of the element.  We conduct two tests to confirm this.  We first inspect the first-order \textit{Cannon} model coefficients which describe the direct quadratic relationship between each spectral pixel and each label.  In Figure A\ref{fig:coeffs}, we plot the first-order \textit{Cannon} coefficients for each label as a function of wavelength.  We mark strong, known lines of each element with a red dashed line and confirm that \textit{The Cannon} is drawing its primary abundance information from those line regions.  Next, in Figure \ref{fig:speccan}, we repeat this exercise by plotting the median spectrum of our sample and highlighting in orange spectral regions that correspond to the strongest 1\% of first-order coefficients for a selection of five elements.  We mark known strong lines of each element with a thick orange line.  These two tests make evident that the primary abundance information retrieved \textit{The Cannon} is coming from strong lines of each element, though it is also clear that \textit{The Cannon} is leveraging the full spectrum to extract abundance information.  This is by design and allows \textit{The Cannon} to achieve its enhanced precision.

\subsubsection{Abundance Uncertainties from \textit{The Cannon} }\label{subsubsec: uncerts}

\begin{figure}
    \centering
    \includegraphics[width=.5\textwidth]{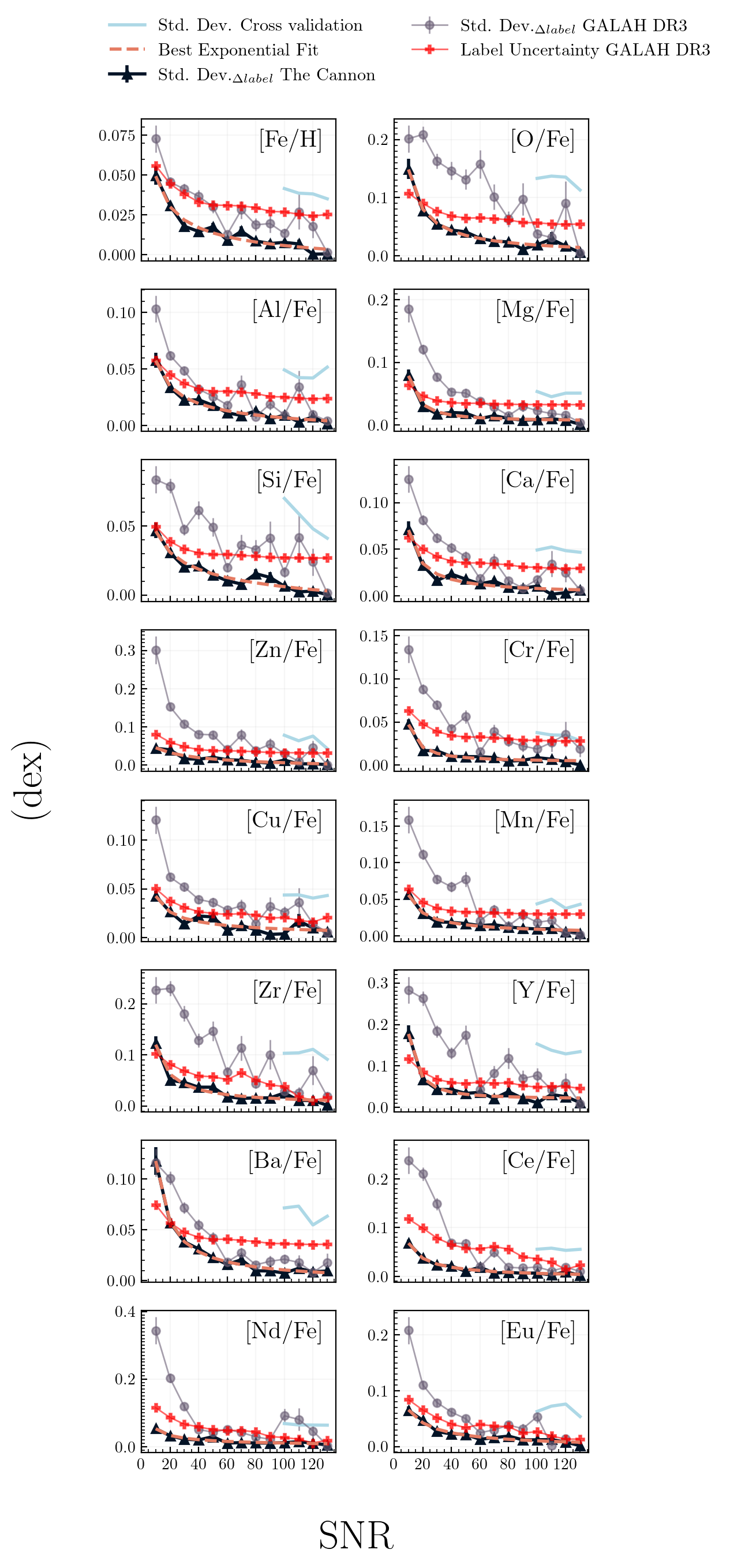}
    \caption{Results of our repeat visit spectrum investigation into the model's precision for each of the 16 elemental abundance ratios that we re-derive in this work.  Solid black triangles present the standard deviation in the difference between the reported \textit{Cannon} label between low-SNR spectra and the highest SNR spectrum as a function of SNR for 387 objects with repeat observations.  An exponential fit to this relationship, which is our final adopted precision, is shown as a dashed orange line.  Gray circles represent the equivalent of the black triangles except for the GALAH-reported abundances.  Red crosses display the mean uncertainty as a function of SNR.  We mark the standard deviation from our cross-validation test (Figure \ref{fig:cannon_performance}) in light blue.  This figure illustrates the enhanced precision achieved by \textit{The Cannon}, highlighted by the difference in the red and black curves.
    }
    \label{fig:cannon_snrvssigma}
\end{figure}

\begin{figure}
    \centering
    \includegraphics[width=.45\textwidth]{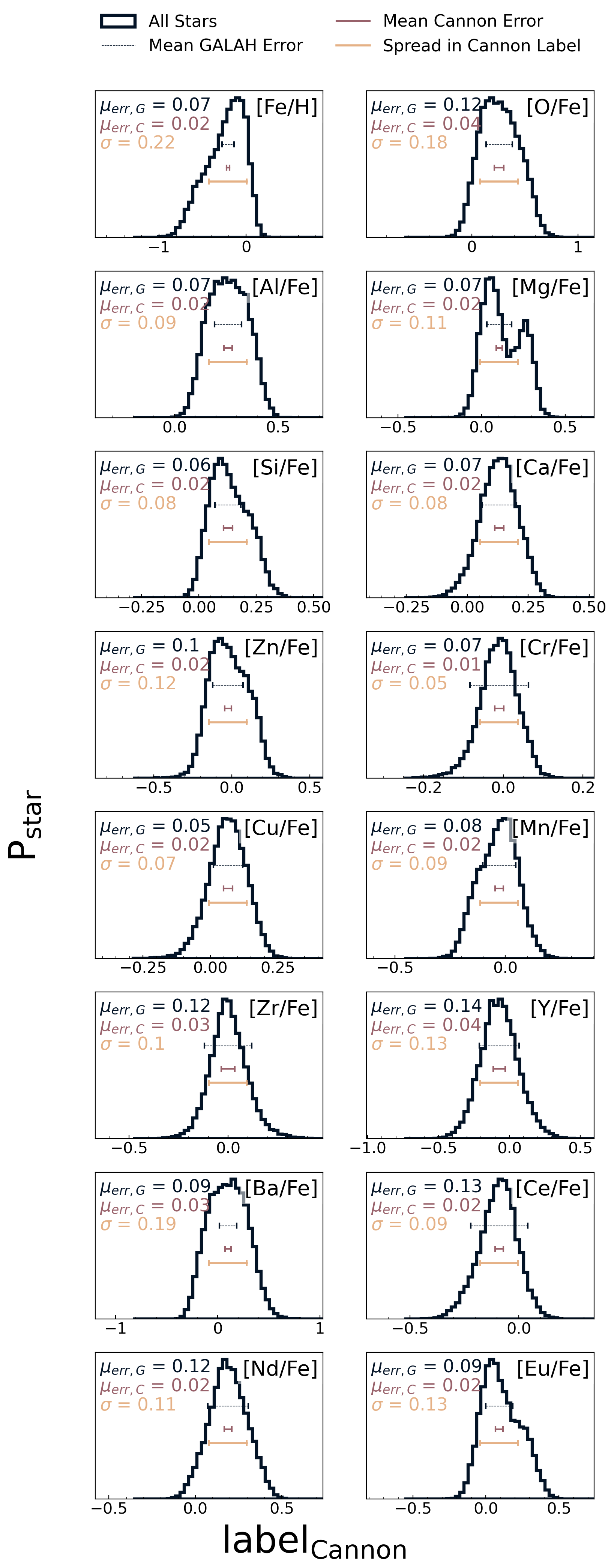}
    \caption{Distributions of the \textit{Cannon}-measured elemental abundances in our final sample.  We mark the mean GALAH-reported error ($\rm \mu_{err, G}$) in each abundance in dark blue, the mean \textit{Cannon}-reported abundance error ($\rm \mu_{err}, C$) in lighter purple, and the spread in the \textit{Cannon}-measured abundances ($\rm \sigma$) in orange  For the majority of elements, \textit{The Cannon} achieves either comparable or lower abundance uncertainties compared to the reported GALAH uncertainties.  Elements with the greatest potential for discriminating power in our sample possess small uncertainty-to-abundance-spread ratios.}
    \label{fig:errspread}
\end{figure}

To determine our final abundance uncertainties, we must take into account two sources of uncertainty.  The first is that reported directly by \textit{The Cannon}, which reflects the dispersion in the final likelihood function for each label.  The second is the external, systematic uncertainty that is best parametrized as a function of spectral SNR \citep[e.g.,][]{Ness2015, Wheeler2020, Nandakumar2022}.  To determine our model's systematic precision, we make use of repeat visit spectra, spectra taken of the same object and later coadded before being measured for the final main GALAH catalog.  Repeat visit spectra present the opportunity to test our model's stability as a function of SNR.  In an ideal case, when running spectra of the same source but with different SNRs through \textit{The Cannon}, our model should always return the same labels regardless of the SNR of the spectrum.  Thus, any change in the model's inferred labels between spectra of the same source but at different SNRs can quantify our SNR dependent label uncertainty.  For this test, we download the GALAH DR3 \code{all\_spec} catalog, which reports stellar parameters for each individual observed spectrum.  We then identify stars with more than one observation by filtering for repeated values in the \code{dr3\_source\_id} column.  We then download 8 nights of data that have significant numbers of targets with repeat-visit spectra (150427, 150428, 150429, 150430, 170912, 170911, 170910, and 170909) and only consider the 387 targets that span the parameter space of our larger data set.  We then produce several instances of each source's spectrum at various SNRs, starting first with single spectra populating the lowest SNR bins, followed by coadded versions of the spectra.  For example, if a source has three total observations, we are able to produce three low, three medium, and one high SNR version of its spectra for the purpose of this experiment.  We then run the spectra through \textit{The Cannon} and, taking the labels reported for the highest SNR spectrum as "truth," measure the dispersion in the difference in the inferred labels between the low- and high-SNR spectra as a function of SNR.  As reported in \citet{Nandakumar2022}, our precision increases with SNR exponentially and plateaus beyond a SNR of 40.  We fit exponential functions to describe the relationship between SNR and label recovery precision and adopt the SNR-dependent dispersions as the external precision of our inferred abundances.

We compute our final label uncertainties by taking the quadratic sum of the internal model uncertainties reported by \textit{The Cannon} and the external uncertainties from our SNR experiment.  In Figure \ref{fig:cannon_snrvssigma}, we compare our repeat-visit abundance dispersion (black triangles) with those of GALAH repeat-visit results (gray circles) and GALAH-reported uncertainties (red crosses) as a function of SNR.  Our resulting abundance precision is improved relative to the GALAH-reported precision by up to a factor of three.  We find that for most elements, the dispersion in the difference between GALAH-reported labels and our \textit{Cannon}-inferred labels from our cross-validation (solid gray line) is similar to the GALAH uncertainty, indicating that the internal precision of \textit{The Cannon} is very high.  In Figure \ref{fig:errspread}, we show the spread in abundance across stars in our sample (black histogram) and compare it to \textit{The Cannon's} mean achieved abundance precision (purple) and that of GALAH DR3 (navy).  Elements such as Zr, Y, and Ce, for example, previously had uncertainties equivalent to the full sample's 1-$\sigma$ abundance spread (orange line) in the element.  Thus, any potential distinguishing power in these elements was thwarted by the GALAH precision.  With our \textit{Cannon}-enhanced abundance precisions, these elements can now be used to potentially distinguish between doppelgangers.

\subsection{Final Catalog}
\begin{figure}
    \centering
    \includegraphics[width=.5\textwidth]{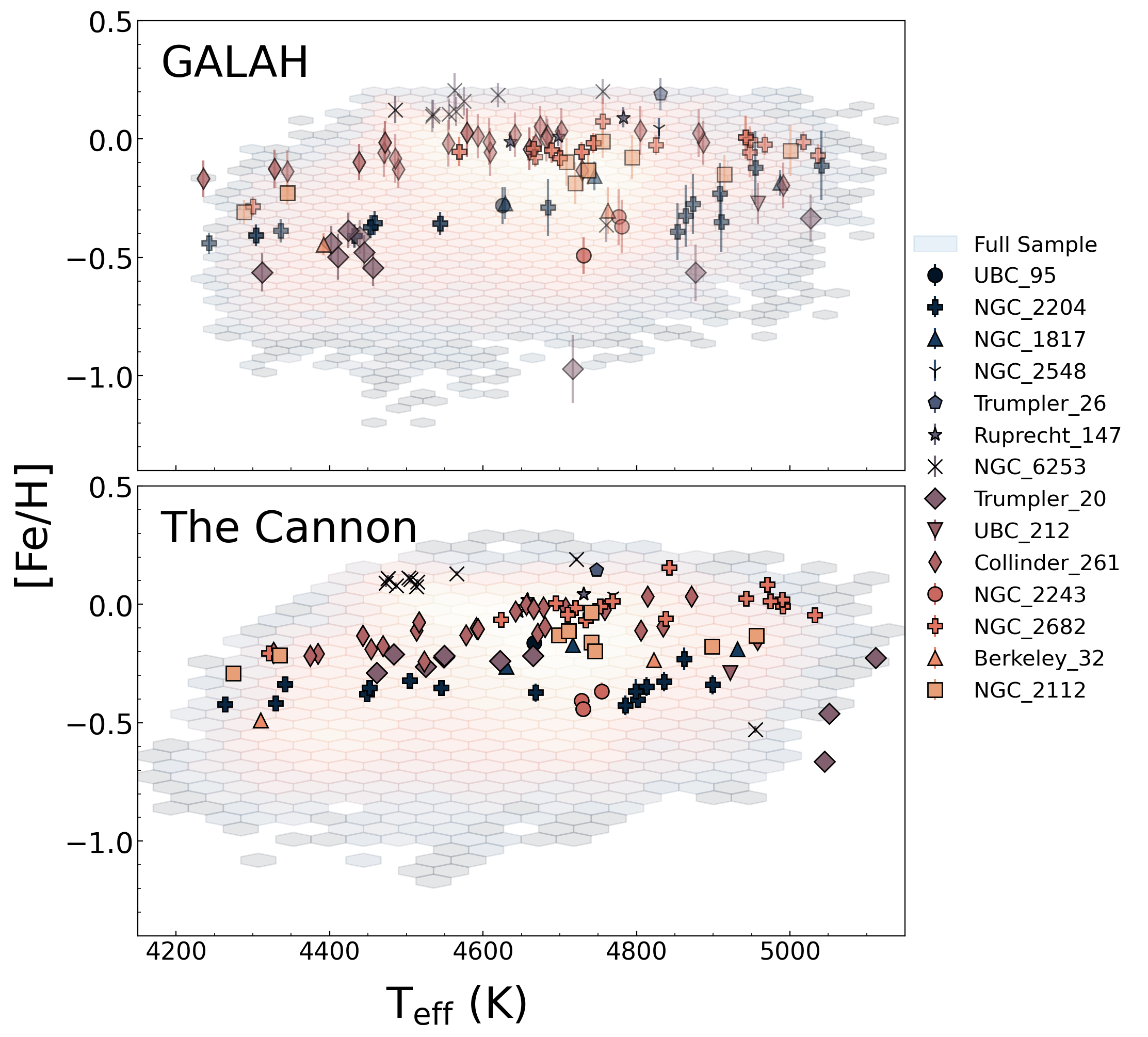}
    \caption{[Fe/H] as a function of $\rm T_{eff}$ for all 122 open cluster stars in our sample. Background shows the full sample's distribution in this plane. Top panel presents the GALAH DR3 abundances and bottom panel presents the \textit{Cannon}   results. The 86 stars in this figure with SNR > 40 and unflagged abundances in all elements (see Section \ref{subsubsec:trainingset}) serve as the reference point for the chemical homogeneity of stars born together, an ingredient in our measurement of the doppelganger rate.}
    \label{fig:cannon_oc_validation}
\end{figure}

Our final catalog consists of 28,120 stars with newly inferred values of \code{v\_broad}, $\rm T_{eff}$, log g, [Fe/H], and [X/Fe] for O, Al, Mg, Si, Ca, Cr, Cu, Zn, Mn, Zr, Y, Ba, Ce, Nd, and Eu for each star that populates the polygon in Figure \ref{fig:training_set}.  In Figure \ref{fig:cannon_oc_validation}, we present the [Fe/H] vs. $\rm T_{eff}$ distributions for the 14 open clusters in our sample, showing the GALAH-reported distribution in the top panel and the \textit{Cannon}-inferred distribution in the bottom panel.  The table schema for our final catalog is included in the Appendix Table A\ref{tab:fin} and the full table is available online.  As mentioned in Section \ref{subsubsec:trainingset}, we flag as unreliable all stars with global $\chi^2$ and individual element $\chi^2$ goodness-of-fit values that exceed two times the degrees of freedom of the spectrum or relevant line mask.  We hereafter only consider stars with unflagged global and individual element abundances.

\begin{figure*}
    \centering
    \includegraphics[width=\textwidth]{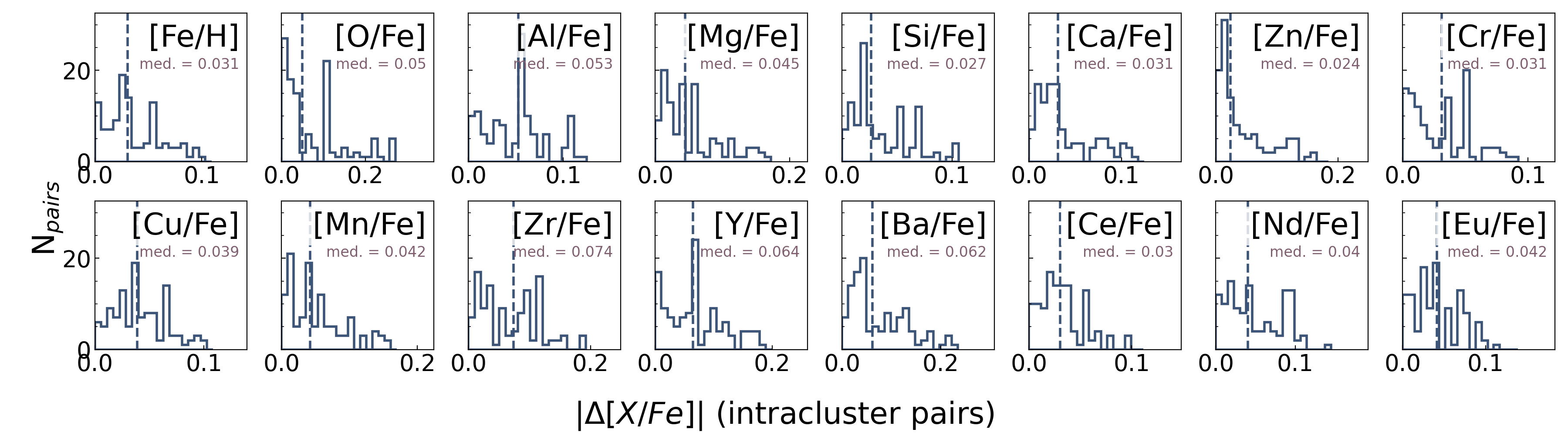}
    \caption{ Distributions of absolute difference in \textit{Cannon}-derived abundance for 122 intracluster pairs with $\rm \Delta T_{eff}$ < 100 K, $\rm \Delta log~g$ < 0.1 in our sample for each element. The median absolute abundance difference value for all pairs is marked with a dashed line and printed in each panel. We see that in general, open clusters in our sample are highly chemically similar but display signs of abundance variation, particularly in the light \textit{s-}process neutron capture elements (Zr, Y, Ba).}
    \label{fig:cannon_chisq_m67}
\end{figure*}

To assess the chemical homogeneity of the open clusters in our sample in light of our re-derived abundances, we draw all possible intracluster pairs with $\Delta \rm T_{eff} < 100K$, $\Delta \rm log g < 0.1~dex$, and unflagged abundances in all elements and present the distributions in absolute difference in abundance for all intracluster pairs in each element in Figure \ref{fig:cannon_chisq_m67}, with median values marked with a dashed line.  Intracluster pairs in general tend to show small abundance differences, with the majority of elements showing median absolute differences in abundance between 0.024 (Zn) and 0.074 (Zr) dex.  It is evident that some intracluster pairs, however, display abundance differences that are large (up to 0.2 dex for Zr, Y, and Ba).

\begin{figure*}
    \centering
    \includegraphics[width=\textwidth]{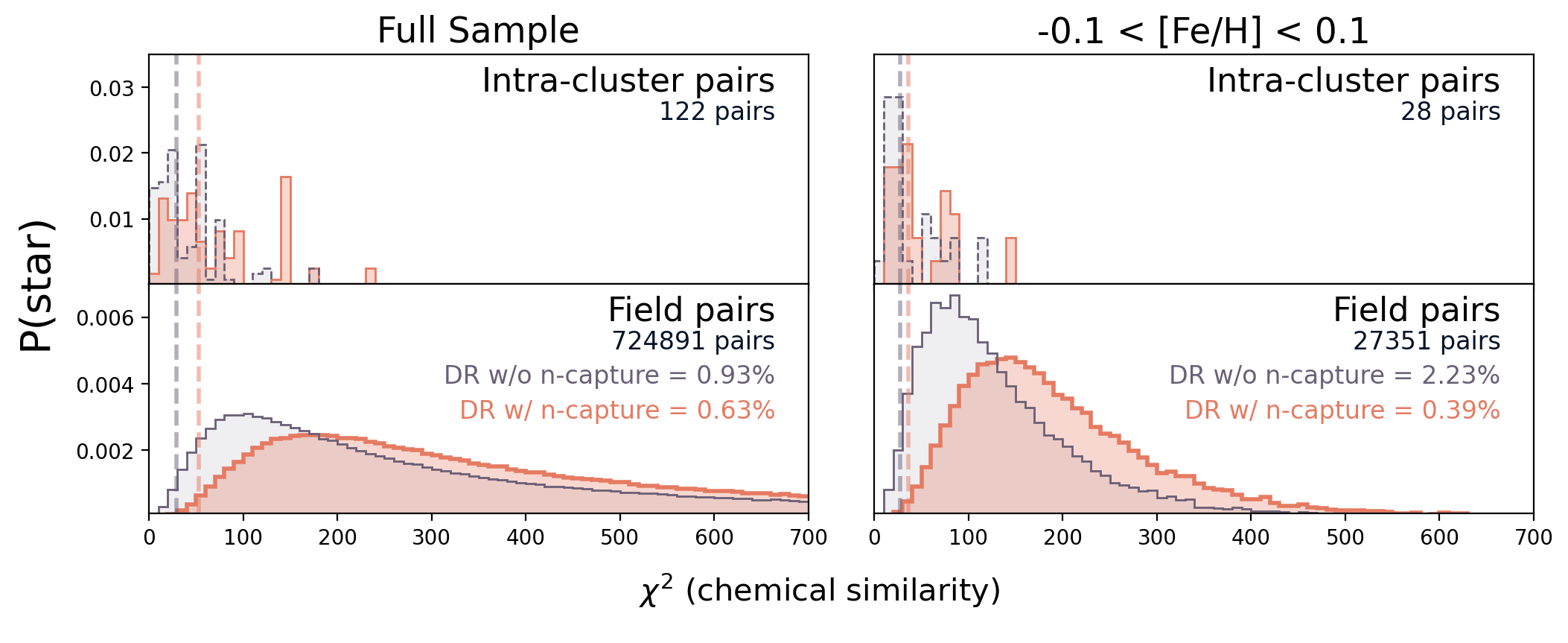}
    \caption{$\chi^2$ (see Equation \ref{eq:chisq}) distributions for intracluster pairs (top panels) and those drawn from the field (bottom panels).  Gray histograms present the $\chi^2$ distributions when considering the light (here, Al), $\alpha$ (Mg, Si, Ca), and iron-peak (Fe, Cr, Cu, Zn, Mn) elements only.  The orange histograms present the $\chi^2$ distributions when considering the aforementioned elements as well as the neutron-capture elements (Zr, Y, Ba, Ce, Nd, Eu).  Vertical lines mark the median $\chi^2$ value for open cluster pairs without (gray) and with (orange) the consideration of neutron-capture elements. Left panels contain the results for our full SNR > 40 sample while the right panels contain results for stars within the -0.1 < [Fe/H] < 0.1 range.  The addition of neutron-capture elements reduces the doppelganger rate by a factor of a third for all field pairs and nearly six for field pairs in the narrow [Fe/H] range.}
    \label{fig:dr}
\end{figure*}

\begin{figure*}
    \centering
    \includegraphics[width=12cm]{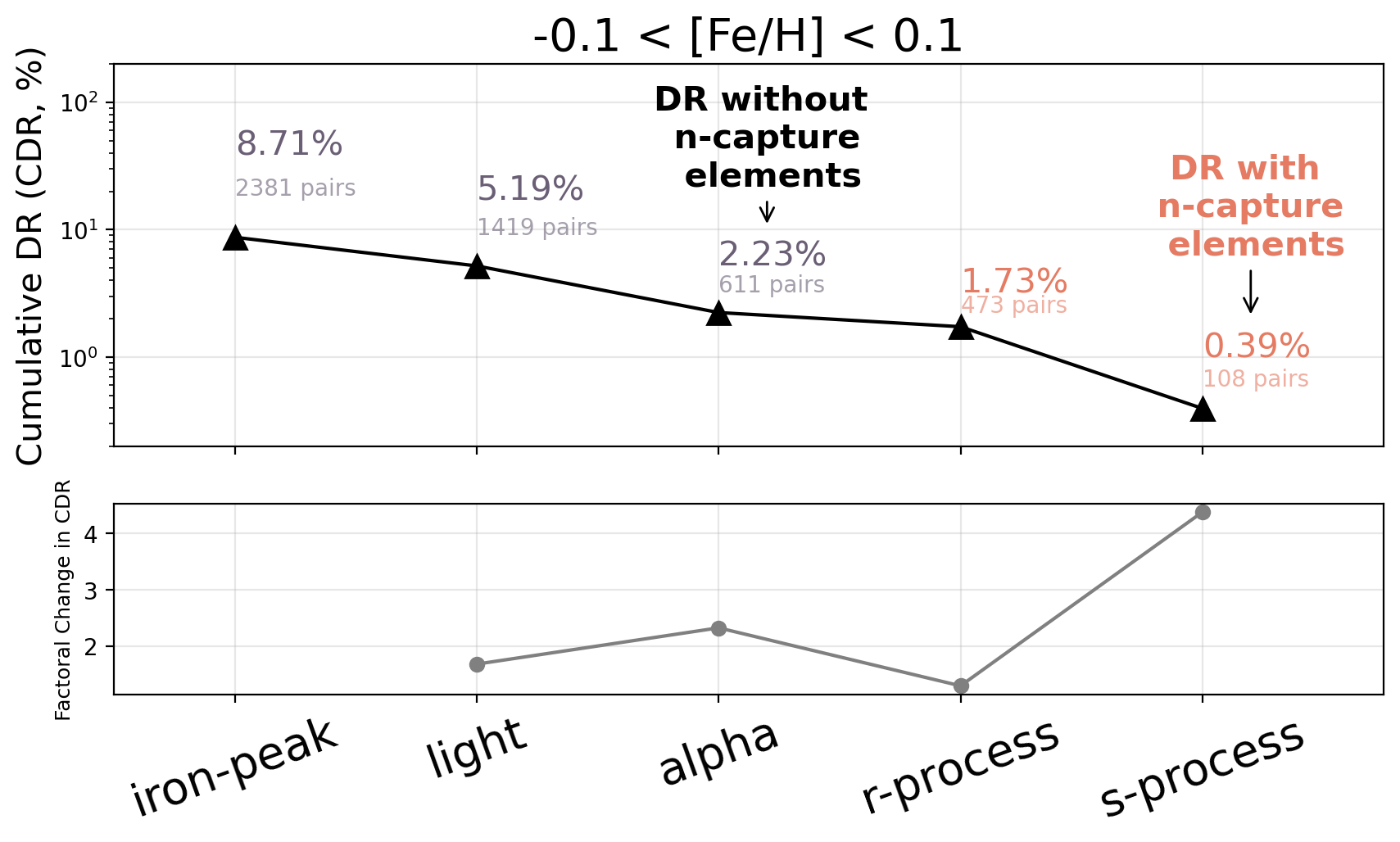}
    \caption{\textit{Top panel:} The cumulative doppelganger rate (CDR) as a function of the addition of each new elemental family.  The CDR represents the total doppelganger rate when considering all elemental elemental families successively. That is, on the far left, we determine the doppelganger rate considering just the iron-peak elements while on the far right, we determine the doppelganger rate with the light, $\alpha$, iron-peak, \textit{r}-, and \textit{s}-process elements. The introduction of neutron-capture elements reduces the doppelganger rate by a factor of 5.75.  It is possible that \textit{s}-process elements have a greater influence on the CDR than the \textit{r}-process elements, but we remind the reader that we only use one \textit{r}-process element, Eu, but five \textit{s}-process elements (Zr, Y, Ba, Ce, and Nd), so the comparison is not straightforward.  \textit{Bottom panel:} The factor with which the CDR changes upon adding each new elemental family.  The y-axis reports the factor with which the doppelganger rate changes as a function of each added elemental family.  The bottom panel illustrates that once pairs are selected by chemical similarity in iron-peak elements, the \textit{s-}process elements possess the greatest additional distinguishing power, followed by the alpha elements, the light elements, and the \textit{r-}process elements.}
    \label{fig:cdr}
\end{figure*}

\begin{figure*}
    \centering
    \includegraphics[width=12cm]{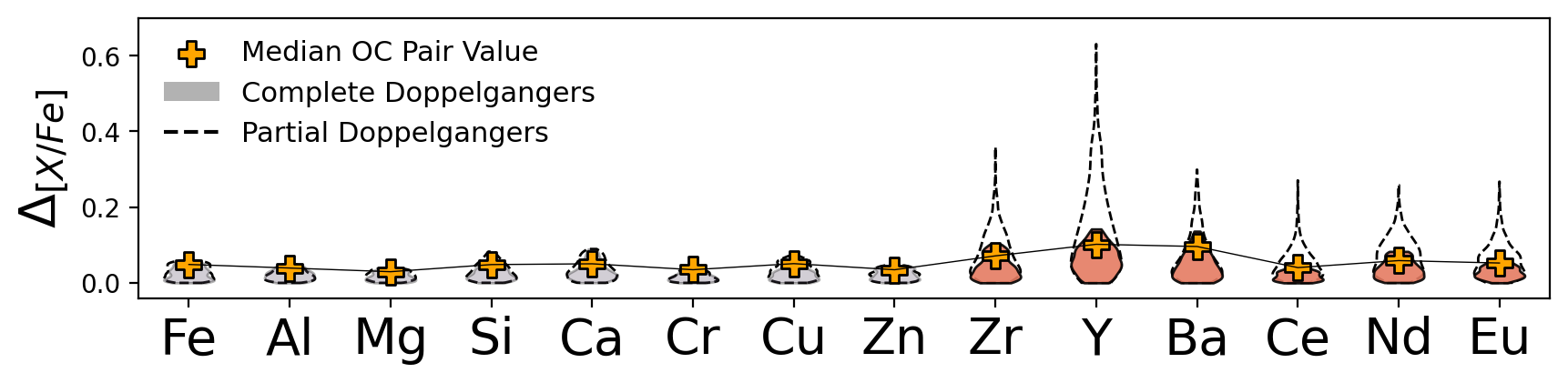}
    \caption{Violin plots showing the absolute difference in abundance ($\rm \Delta$ [Fe/H] for Fe,  $\Delta$ [X/Fe] for the remaining elements) between stars in doppelganger pairs for each element.  The abundance difference distributions for partial doppelgangers (those that are doppelganger exclusively in light, $\alpha$, and iron-peak elements) are represented by the empty violin plots.  The same distributions for complete doppelgangers (doppelganger in all measured elements) are represented by the filled violin plots.  The median abundance difference in each element for intracluster pairs is represented by the orange crosses.  This figure illustrates that random pairs of field stars can appear as chemically similar as open cluster stars in the lighter elements but show strong deviations in the heavier elements.}
    \label{fig:abundviolin}
\end{figure*}

\begin{figure*}
    \centering
    \includegraphics[width=12cm]{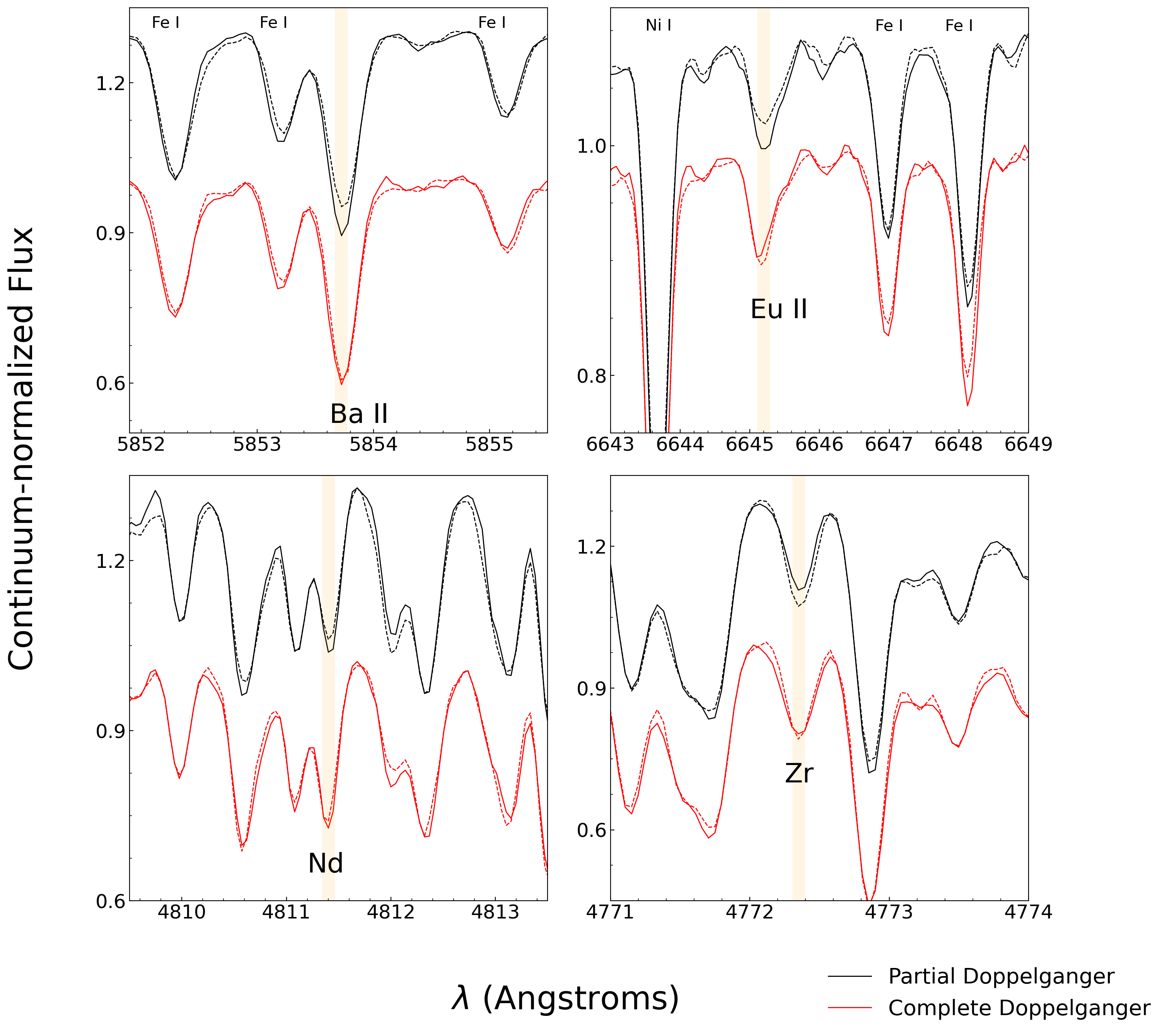}
    \caption{Example spectra of the two types of doppelganger pairs: partial doppelgangers (black) and complete doppelgangers (red).  In each panel, we highlight spectral regions around neutron-capture element lines, with the faint orange line marking the line core of the labelled element.  Note how the complete doppelganger spectra match in the highlighted neutron-capture features while the partial doppelgangers show deviations, as expected given our definitions of these two populations.}
    \label{fig:doppspec}
\end{figure*}

\begin{figure}
    \centering
    \includegraphics[width=.5\textwidth]{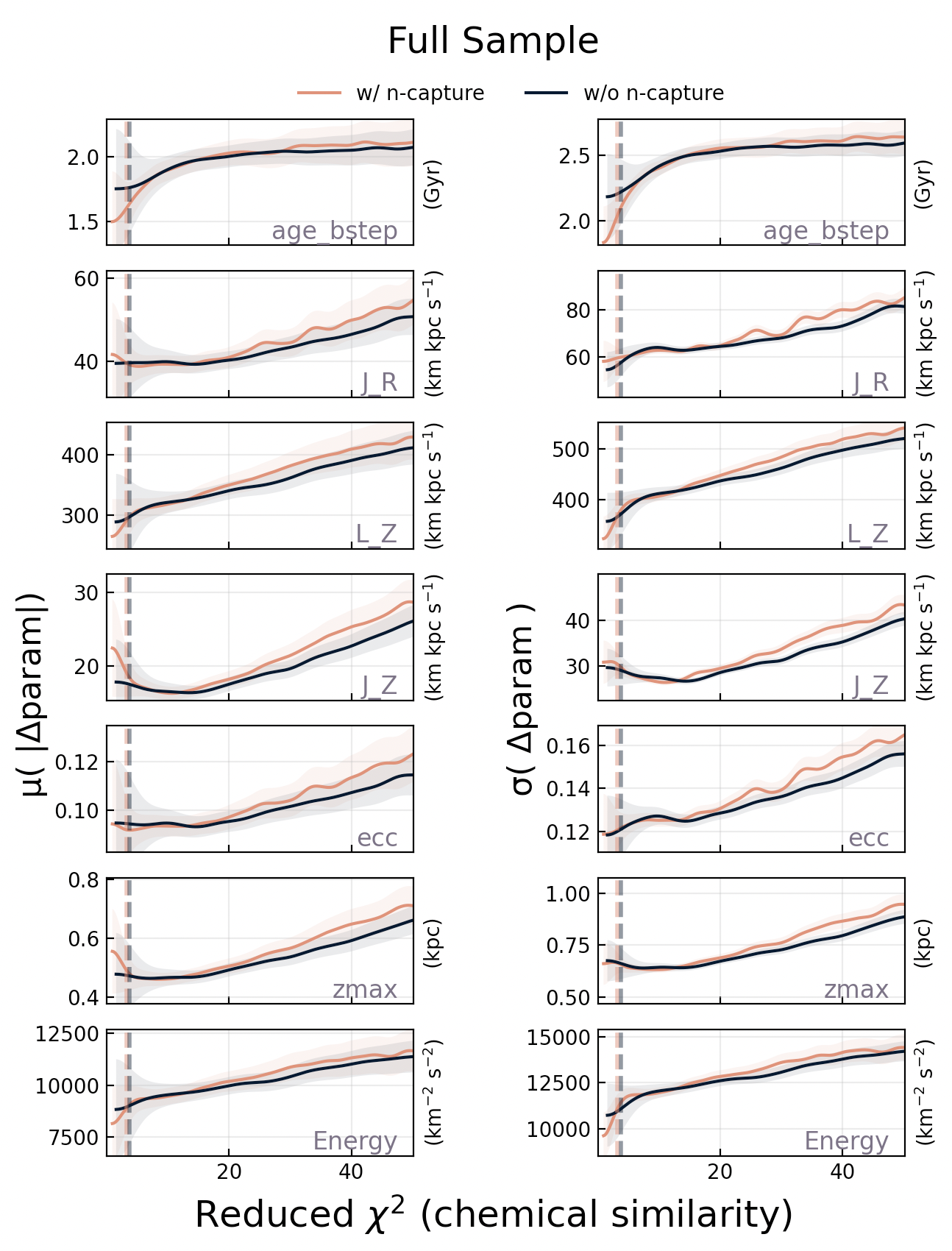}
    \caption{Running mean (left column) and standard deviation (right column) in (absolute, for left column) age and dynamical property difference for random pairs of field stars as a function of their reduced $\chi^2$, a measure of their chemical similarity normalized by the number of elements considered (i.e., the degrees of freedom).  The gray curve omits the neutron-capture elements while the orange curve includes the neutron-capture elements, and the thick dashed line represents the median reduced $\chi^2$ for intracluster pairs.  While chemical similarity globally correlates with age or dynamical similarity, complete doppelgangers (orange curve that lies leftward of the dashed line) do not appear to be more similar in age and dynamical parameters than partial doppelgangers (gray curve that lies leftward of dashed line) using our defined $\chi^2$ metric.}
    \label{fig:delta_age}
\end{figure}

\subsection{Measuring the Doppelganger Rate in GALAH DR3}\label{subsec:id_dopps}
Instead of measuring the doppelganger rate in our full sample, we measure it in a higher quality subset of our data.  This is because the doppelganger rate is sensitive to abundance precision and our choice of open cluster reference pairs.  In general, in the low SNR regime, where abundance uncertainties are high, pairs of stars will tend to look more chemically similar within uncertainties.  We find that \code{snr\_c2\_iraf} > 40 is an ideal SNR lower limit for measuring a meaningful doppelganger rate in our sample as it ensures we maximize our abundance precision.  This SNR cutoff enables us to sample 47 stars across five clusters (NGC 2112, NGC 6253, NGC 2204, Collinder 261, and M 67) and 13,375 stars in the field.

To build our reference sample of open cluster stars, we draw all possible combinations of intracluster pairs where stars in the pair have $\rm \Delta T_{eff}$ < 100 K and $\rm \Delta log~g$ < 0.1 dex.  The $\rm \Delta T_{eff}$ and $\rm \Delta log~g$ requirement ensures that we avoid potential systematic trends between abundance and $\rm T_{eff}, log~g$ to artificially enhance or minimize the abundance similarity of stars in a pair.  We ultimately build a population of 122 intracluster pairs that serve as a reference point for the chemical similarity of stars born together in open clusters.  For our field sample, we draw one million unique pairs of stars that are not members of \citet{Spina2021} open clusters and that satisfy the same $\rm \Delta T_{eff}$, $\rm \Delta log~g$ requirements as those for open cluster pairs.  These field pairs serve to sample the chemical diversity (or lack thereof) of the phase-mixed Galactic disk population.

After building our intracluster and field pair samples, we measure the doppelganger rate using the method of \citet{Ness2018}, computing a global $\chi^2$ value using Equation \ref{eq:chisq} for each pair of stars and defining doppelgangers to be field pairs with $\chi^2$ less than the median $\chi^2$ value of intracluster pairs in the considered elements.

\section{Results}\label{sec:results}
\subsection{Impact of neutron-capture elements on the DR}
Our primary question in this work asks whether the neutron-capture elements affect the doppelganger rate.  In Figure \ref{fig:dr}, we use Equation \ref{eq:chisq} to determine the $\chi^2$ distributions, a proxy for chemical similarity, of intracluster pairs (top panels) and field pairs (bottom panels).  We compare the $\chi^2$ distribution of the field pairs with that of the intracluster pairs when excluding (gray) and including (orange) the neutron-capture elements.  Faint dashed lines correspond to the median $\chi^2$ value for intracluster pairs when excluding (gray) and including (orange) neutron-capture elements. The left panel presents the aforementioned for the full SNR > 40 sample, while the right panel presents it for a subset of stars with -0.1 < [Fe/H] < 0.1.  As found in \citet{Ness2018}, stars born together in open clusters are far more chemically similar than random field pairs.  However, there exist field star pairs that are just as, if not more, chemically alike than intracluster pairs, as quantitatively shown in their $\chi^2$ values. These field pairs are deemed doppelgangers.  As expected, adding neutron-capture elements (Zr, Y, Ba, Ce, Nd, and Eu) shifts the $\chi^2$ distributions for each population due to adding degrees of freedom to the $\chi^2$ calculation.  However, the $\chi^2$ distribution of open cluster pairs shifts comparatively \textit{less} than that of random field pairs.  If neutron-capture elements had no further distinguishing power compared to the lighter elements, then the doppelganger rate would remain constant when including them.  However, including neutron-capture elements decreases the doppelganger rate from 0.9\% to 0.6\% in our full sample and from 2.2\% to 0.4\% in our sample with a narrow [Fe/H] range. Figure \ref{fig:dr} illustrates that the neutron-capture elements have a subtle but non-negligible affect on the doppelganger rate.  

\subsection{Which elements matter most in distinguishing disk stars?}
In Figure \ref{fig:cdr}, we present the impact of each successive elemental family on the doppelganger rate via a \textit{cumulative} doppelganger rate (CDR) computed in the narrower -0.1 < [Fe/H] < 0.1 range.  In the top panel, we report the measured CDR as a function of each elemental family and the elemental families lighter than it.  For example, when we plot the CDR associated with the iron-peak elements, we plot the percentage of pairs that have $\chi^2$ values computed using Fe, Cr, Cu, Mn, and Zn  less than the median of intracluster pairs.  Next, for the light elements, we plot the same as for the iron-peak elements but this time also considering Al in our $\chi^2$ calculations.  Next, for $\alpha$ elements, we plot the same but considering Al, Fe, Cr, Cu, Mn, Zn, Mg, Ca, and Si.  This continues on as one moves rightward, ultimately terminating with the \textit{s-}process elements.  In the bottom panel, we illustrate the practical discriminating power of each elemental family by reporting the multiplicative factor with which the CDR changes upon the addition of each new family.  We find that once doppelgangers are identified using the iron-peak elements, the \textit{s-}process elements possess the greatest additional distinguishing power, followed by the $\alpha-$elements, the light elements, and the \textit{r-}process elements.  We note that this narrow [Fe/H] range contains primarily low-$\alpha$ disk stars, and this may play a role in the relatively low distinguishing power of the $\alpha-$elements. As in Figure \ref{fig:dr}, we find that the doppelganger rate considering light, $\alpha$, and iron-peak elements is 2.23\% for stars with -0.1 < [Fe/H] < 0.1.  With the addition of the neutron-capture elements (Eu, Zr, Y, Ba, Ce, and Nd), the doppelganger rate reduces to 0.39\%, by a factor of 5.75 (with \textit{r}-process element Eu reducing it to 1.73\% and \textit{s}-process elements reducing it to the final 0.39\%.)

\subsection{Partial vs. Complete Doppelgangers}
Upon discovering that neutron-capture elements affect the doppelganger rate, we isolate pairs of field stars that are \textit{partial} doppelgangers (that is, they satisfy doppelganger requirements in the light, $\alpha$, and iron-peak elements) from pairs of field stars that are \textit{complete} doppelgangers (that is, they satisfy doppelganger requirements in all elements).  In Figure \ref{fig:abundviolin}, we show the distributions in absolute difference in abundance for partial (open violin plots) and complete (filled violin plots) doppelgangers and also include as reference the median absolute difference in abundance for intracluster pairs (yellow crosses).  We note that this plot reports abundance differences between stars in a pair, not $\chi^2$ values, which we use in the actual computation of the doppelganger rate.  This figure illustrates that there exist pairs of field stars that are as chemically similar in the light, $\alpha$, and iron-peak elements as stars born together but deviate in the neutron-capture elements by up to 0.6 dex.  In Figure \ref{fig:doppspec}, we show example spectra of partial doppelgangers (black) and complete doppelgangers (red).  Note that despite both pairs of stars satisfying doppelganger requirements in the lighter elements, the partial doppelganger spectra show deviations in the neutron-capture lines (highlighted in faint orange) while the complete doppelgangers do not.  We note here that since neutron-capture elements can distinguish partial from complete doppelgangers, the results of \citet[][]{Ness2018} may overestimate the doppelganger rate due to the exclusion of this elemental family. We explore this possibility in Manea et al., in prep..

It is of physical interest to investigate why some pairs are partial doppelgangers while others are complete doppelgangers.  Adopting the dynamical parameters and age estimates from the associated GALAH value-added catalogs (see Section \ref{sec:data}), we compare the similarity in these characteristics for pairs of stars that are partial versus complete doppelgangers. In Figure \ref{fig:delta_age}, we plot the running mean (left column) and standard deviation (right column) in the (absolute for the left column) difference in age (top panel) and a series of dynamical characteristics ($\rm J_R$, $\rm L_Z$, $\rm J_Z$, eccentricity,  $\rm z_{max}$, and energy) for pairs as a function of their chemical similarity, which we measure using a reduced $\chi^2$, defined by the formula in Equation \ref{eq:chisq} but further divided by the number of degrees of freedom (e.g., number of elements considered, which is 9 when ignoring the neutron-capture elements, marked by the gray curves, and 15 when including them, marked by the orange curves.)  Dashed lines represent the median reduced $\chi^2$ value for intracluster pairs.  Shading represents the uncertainty, which for the left panels is +/-~$\sigma$/(N)$^{1/2}$ and for the right panel is +/-~$\sigma$/(2N)$^{1/2}$ where N is the number of stars in the bin. $\chi^2$ correlates strongly with similarity in age or dynamical characteristics.  However, we do not find clear evidence that complete doppelgangers (orange curve leftward of the dashed lines) possess more similar age and dynamical parameters relative to partial doppelgangers (gray curve leftward of dashed lines) using the $\chi^2$ metric.  However, these results should be followed up using higher precision ages and dynamical characteristics, as any potential signature may be concealed beneath large uncertainties in age and orbital properties.  Furthermore, this may be due to our definition of $\chi^2$, which is a sum of chemical similarity in all elements.  Treating each element individually may yield different results, but we leave this for a future investigation.

\section{Discussion}\label{sec:discussion}
Our investigations into the doppelganger rate in GALAH DR3 conclude that the neutron-capture elements possess subtle but non-negligible distinguishing power in the Milky Way disk that is important to consider when chemically characterizing stars.  As captured in Figures \ref{fig:dr} and \ref{fig:cdr}, these results suggest that the neutron-capture elements contain information distinct from that of the lighter elements and thus add to the dimensionality of Milky Way abundance space.  For example, when ignoring the neutron-capture elements, we identify over 6,700 pairs of apparently-unrelated field stars that are as chemically similar as stars in open clusters.  However, when introducing the neutron-capture elements, we find that 1/3rd of those pairs are not in fact as chemically similar as stars in open clusters when considering the neutron-capture elements.  When restricting ourselves to a narrow [Fe/H] range (-0.1 < [Fe/H] < 0.1), we identify over 600 pairs of doppelgangers when omitting the neutron-capture elements, and upon introducing the neutron-capture elements, we are left with just $\sim$108 pairs (a reduction by a factor of 5.75).  These results directly support those of \citet{Griffith2022}, which found that introducing a chemical dimension corresponding to AGB star nucleosynthesis is required to reproduce the chemical distribution of the GALAH survey.  Additionally, these results support those of \citet{Lambert2016}, which found that open clusters deviate most significantly in the neutron-capture elements even when sharing light (Z<30) element compositions, suggesting that neutron-capture elements contain additional distinguishing power not captured by the lighter elements.

\subsection{Possible Physical Origins of Enhanced Distinguishing Power of Neutron-capture Elements}
There are potential physical explanations for why neutron-capture elements could add to the chemical dimensionality of Milky Way stars.  Their unique production and dispersal sites may embed neutron-capture elemental abundances with temporal and spatial information that is distinct from that contained in the lighter (Z<30) elements.  For example, several works have identified that \textit{s}-process element abundances, when compared to $\alpha$-element abundances, can be effective chemical clocks that probe stellar age both within and outside the Milky Way \citep[e.g.,][]{Feltzing2017, Skuladottier2019, Ratcliffe2023}.  This is due to their primary nucleosynthetic origins in AGB stars, which are lower mass ($\rm M < 8-10 M_{\odot}$) than the high mass ($\rm M > 10 M_{\odot}$) stars that produce the $\alpha$ elements \citep[e.g.,][]{Karakas2014}.  Because stellar mass strongly influences stellar lifetime, \textit{s}-process elements have a longer delay time for enrichment in the Galaxy with respect to the $\alpha$-elements.  When stars form from gas that is enriched in products of both AGB star and core collapse supernova nucleosynthesis, this delay-time difference enables \textit{s}-process elements to trace the star's age when compared to $\alpha$ elements, though the relationship between the [\textit{s}-process/alpha] ratio and stellar age is neither universal nor simple \citep[e.g.,][]{Casali2020}.  In addition to age information, \textit{s}-process elements may also add information about stellar birth position that is not captured by supernova-produced elements.  Simulations suggest that the \textit{s}-process elements have shorter correlation lengths in the Milky Way interstellar medium relative to the lighter elements due to the nature of their dispersal \citep[e.g.,][]{Armillotta2018, Krumholz2018, Emerick2020}. The \textit{s}-process elements are dispersed via relatively gentle, localized AGB star winds, contrary to the highly energetic dispersal of the lighter elements via supernovae \citep[e.g.,][]{Cox2012}.  The more localized nature of their dispersal may thus allow for greater variation in \textit{s}-process composition as a function of location within the interstellar medium.  Thus, stars born together in pockets of this enriched gas will be chemically similar in these heavier elements, but stars across different birth groups, even at fixed metallicity, may differ in their  \textit{s}-process composition.  We indeed see evidence of this when studying open clusters \citep[e.g.,][]{Lambert2016}.  This could imply that either a) the timescale between the enrichment of \textit{s}-process products into the interstellar medium and the formation of stars from this material is shorter than the timescale with which the interstellar medium mixes away gas-phase abundance variations, or b) the interstellar medium is less efficient at mixing \textit{s-}process products (again potentially due to their less-energetic dispersal compared to supernova products, \citealp{Krumholz2018, Emerick2020}).

The \textit{r}-process elements also have physical reasons to add a unique dimension to Milky Way abundance space.  The \textit{r}-process elements are believed to be formed in stochastic events such as magnetorotational supernovae \citep[e.g.,][]{Siegel2017, Halevi2018, Siegel2019} and compact object mergers \citep[e.g.,][]{Korobkin2012}. The stochastic nature of their synthesis could allow them to carry additional information about stellar birth position or age.  We note that the origin of the \textit{r}-process elements is still a major open question in the field \citep[e.g.,][]{Lian2023, Kobayashi2023}.

The results of our analysis of the CDR (Figure \ref{fig:cdr}) appear to suggest that the \textit{r}-process elements hold less distinguishing power than the \textit{s}-process elements.  We caution that Eu is the only pure \textit{r}-process element in our analysis that represents this nucleosynthetic family, whereas we have five elements representing the \textit{s}-process family, one of which also has significant \textit{r}-process contribution (Nd, e.g., \citealp{Kobayashi2020}).  It is thus possible that much of the distinguishing power of \textit{r}-process elements is overshadowed by the numerous \textit{s}-process elements we sample.  However, the \textit{s}-process elements possessing greater distinguishing power than the \textit{r}-process elements also has physical support.  Leading potential origins of \textit{r}-process elements involve dispersal via energetic supernovae, and this may cause them to behave differently than \textit{s}-process elements in the ISM.  Further work into the dependence of the doppelganger rate on \textit{r}-process elements must be conducted to clarify whether they truly carry less distinguishing power than \textit{s}-process elements, or if this is just a consequence of the choice of elements considered.

Given the neutron-capture elements that we consider in this work, the doppelganger rate reduces from $\sim$2.2\% to 0.4\%. Whilst this is a substantial drop in the doppelganger rate, and informative for nucleosynthetic sources and mixing, this still prohibitive for the prospect of strong chemical tagging in the Milky Way Galaxy.  This rate, which measures the probability of which random stars are chemically as similar as stars born together, is still a factor of at $\approx$1000-10000 times greater than the expected rate of recovering true birth pairs in a disk. This is assuming that clusters form from with a typical mass of $1 \times 10^4\,\mathrm{M_\odot}$ - $1 \times 10^6\,\mathrm{M_\odot}$ \citep[e.g.,][]{BlandHawthorn2010}.

\subsection{Analysis Limitations and Looking Ahead to GALAH DR4}
We emphasize that the doppelganger rate is influenced by several factors, some intrinsic to the Milky Way (e.g., mixing efficiency in the interstellar medium, variations in nucleosynthetic yields) and some caused by the available data.  The results of this work are limited by the precision of derived elemental abundances.  With increased abundance precision, we may find that the doppelganger rate decreases even further.  However, the results of this work suggest that at the abundance precision of GALAH DR3 combined with \textit{The Cannon}, we are able to harness the distinguishing power of neutron-capture elements.  We note that if we repeat this experiment using the provided GALAH DR3 abundances of a high SNR (SNR > 100), high precision subsample of the catalog, we obtain the same qualitative results.  This work is also limited by the number of elements that we consider in our analysis.  For the purpose of building a reasonably sizeable training set, we could not consider all $\sim$30 elements reported by GALAH, as it is rare for stars to have high-fidelity abundance measurements in all elements.  As such, we only considered between 1 and 5 elements from each nucleosynthetic family.

The upcoming fourth data release of GALAH is an ideal environment for expanding on this experiment.  GALAH DR4 will have a larger sample size overall and enhanced sampling of open cluster stars.  Enhancing the reference open cluster sample would allow for a more granular exploration of the doppelganger rate as a function of metallicity.  In this work, we are limited by the small number of open cluster stars sampled by GALAH DR3, and separating the data into metallicity bins would make the number of reference open cluster stars per metallicity bin prohibitively small. In DR4, the number of sampled open cluster stars will be doubled, and the sampled open clusters will span -2 < [Fe/H] < 1, where as DR3 only reasonably samples open clusters -0.5 < [Fe/H] < 0.5. Furthermore, GALAH DR4 will have improved abundance precision, further minimizing its extrinsic effect on the measured doppelganger rate.  Finally, GALAH DR4 will provide a larger set of stars sampled in a wider range of elements, enabling the consideration of a greater number of elements and thus allowing for a result that better reflects the intrinsic doppelganger rate of our Galaxy.

\section{Conclusions}\label{sec:conclusion}
In this work, we measure the doppelganger rate among red clump and red giant stars in GALAH DR3.  The doppelganger rate measures the rate at which randomly drawn pairs of apparently unrelated field stars appear to be as chemically similar as stars born together.  It probes the chemical diversity of Milky Way stars, the chemical dimensionality of Milky Way abundance space, and the complexity with which Galactic chemical evolution operates.  After re-deriving stellar parameters and abundances with \textit{The Cannon}, we measure the chemical doppelganger rate.  We find that 0.9\% of random pairs of fields stars are doppelgangers in the light-, $\alpha$-, and iron-peak elements.  This number increases to $\sim$2.2\% when we restrict ourselves to stars in the -0.1 < [Fe/H] < 0.1 dex range.  However, we find that including neutron-capture elements Zr, Y, Ba, Ce, Nd, and Eu in our analysis decreases the doppelganger rate significantly.  When considering our full sample, the neutron-capture elements reduce the doppelganger rate to 0.6\%, and when restricting to the -0.1 < [Fe/H] < 0.1 dex range, the doppelganger rate drops to 0.4\%, by nearly a factor of 6 relative to the rate measured considering only the lighter (Z < 30) elements.  In other words, up to 85\% of stars that are highly chemically similar in the lighter elements deviate in their neutron-capture element abundances.  Chemical similarity strongly correlates with similarity in age or dynamics.  However, we do not identify any clear signatures that complete doppelgangers, pairs of stars that are doppelganger in the light, $\alpha$, iron-peak, \textit{and} neutron-capture elements, are more similar in age and dynamical characteristics than partial doppelgangers, those that are doppelganger in the lighter elements but show deviations in the neutron-capture elements.  However, these results are not conclusive, so additional follow up work should be done to to further explore this.  Finally, our results suggest that the \textit{s-}process elements may carry greater distinguishing power in our sample than the \textit{r-}process elements, though we urge additional follow-up to confirm this.

This work highlights that the neutron-capture elements carry unique information that is distinct from that found in the light-, $\alpha$-, and iron-peak elements and are thus important tools in the chemical characterization of Milky Way stars.  Despite their enhanced distinguishing power, our final doppelganger rate of $\sim$0.4\% suggests that neutron-capture elements measured at the precision of this work are likely not sufficient to satisfy the requirements for strong chemical tagging.  However, our results illustrate that neutron-capture elements can distinguish between 85\% of stars that appear chemically similar in the light, $\alpha$, and iron-peak elements, suggesting that these heavy elements, particularly the \textit{s-}process elements, are potentially important tools for the weak chemical tagging of stars to known clusters and stellar populations.  Our results motivate the need for continued work to improve atomic data for the heavy elements and enhance our ability to extract precise and accurate neutron-capture elemental abundances from stellar spectra.

\section*{Acknowledgements}
CM is supported through the University of Texas at Austin Graduate Continuing Fellowship.  KH  acknowledge support from the National Science Foundation grant AST-1907417 and AST-2108736 and from the Wootton Center for Astrophysical Plasma Properties funded under the United States Department of Energy collaborative agreement DE-NA0003843. This work was performed in part at the Aspen Center for Physics, which is supported by National Science Foundation grant PHY-1607611. This work was also performed in part at the Simons Foundation Flatiron Institute's Center for Computational Astrophysics during KH's tenure as an IDEA Fellow. This work was supported by the Australian Research Council Centre of Excellence for All Sky Astrophysics in 3 Dimensions (ASTRO 3D), through project number CE170100013. DBZ and SLM acknowledge the support of the Australian Research Council through Discovery Project grant DP220102254, and SLM acknowledges the support of the UNSW Scientia Fellowship Program.

The following software and programming languages made this research possible: topcat (Python (version 3.9) and its packages astropy (version 2.0; \citealt{Astropy}), scipy \citep{scipy}, matplotlib \citep{matplotlib}, pandas (version 0.20.2; \citealt{pandas}) and  NumPy \citep{numpy}. This research has made use of the VizieR catalog access tool, CDS, Strasbourg, France. The original description of the VizieR service was published in A\&AS 143, 23.  Colour maps used in figures were adopted from those created by Fabio Crameri (\url{http://doi.org/10.5281/zenodo.1243862}).

\section*{Data Availability}
This work has made use of data from the European Space Agency (ESA) mission Gaia (\url{https://www.cosmos. esa.int/gaia}), processed by the Gaia Data Processing and Analysis Consortium (DPAC, \url{https://www.cosmos.esa.int/web/gaia/dpac/consortium}). Funding for the DPAC has been provided by national institutions, in particular the institutions participating in the Gaia Multilateral Agreement.

This work has also made use of GALAH DR3, based on data acquired through the Australian Astronomical Observatory, under programmes: A/2013B/13 (The GALAH pilot survey); A/2014A/25, A/2015A/19, A2017A/18 (The GALAH survey). We acknowledge the traditional owners of the land on which the AAT stands, the Gamilaraay people, and pay our respects to elders past and present.  The GALAH DR3 data underlying this work are available in the Data Central at \url{https://cloud.datacentral.org.au/teamdata/ GALAH/public/GALAH_DR3/} and can be accessed with the unique identifier \code{galah$\_$dr3} for this release and \code{sobject$\_$id} for each spectrum.


\bibliography{main} 

\begin{thebibliography}{}
\expandafter\ifx\csname natexlab\endcsname\relax\def\natexlab#1{#1}\fi
\providecommand{\url}[1]{\href{#1}{#1}}
\providecommand{\dodoi}[1]{doi:~\href{http://doi.org/#1}{\nolinkurl{#1}}}
\providecommand{\doeprint}[1]{\href{http://ascl.net/#1}{\nolinkurl{http://ascl.net/#1}}}
\providecommand{\doarXiv}[1]{\href{https://arxiv.org/abs/#1}{\nolinkurl{https://arxiv.org/abs/#1}}}

\bibitem[{{Abdurro'uf} {et~al.}(2022){Abdurro'uf}, {Accetta}, {Aerts}, {Silva
  Aguirre}, {Ahumada}, {Ajgaonkar}, {Filiz Ak}, {Alam}, {Allende Prieto},
  {Almeida}, {Anders}, {Anderson}, {Andrews}, {Anguiano}, {Aquino-Ort{\'\i}z},
  {Arag{\'o}n-Salamanca}, {Argudo-Fern{\'a}ndez}, {Ata}, {Aubert},
  {Avila-Reese}, {Badenes}, {Barb{\'a}}, {Barger}, {Barrera-Ballesteros},
  {Beaton}, {Beers}, {Belfiore}, {Bender}, {Bernardi}, {Bershady}, {Beutler},
  {Bidin}, {Bird}, {Bizyaev}, {Blanc}, {Blanton}, {Boardman}, {Bolton},
  {Boquien}, {Borissova}, {Bovy}, {Brandt}, {Brown}, {Brownstein}, {Brusa},
  {Buchner}, {Bundy}, {Burchett}, {Bureau}, {Burgasser}, {Cabang}, {Campbell},
  {Cappellari}, {Carlberg}, {Wanderley}, {Carrera}, {Cash}, {Chen}, {Chen},
  {Cherinka}, {Chiappini}, {Choi}, {Chojnowski}, {Chung}, {Clerc}, {Cohen},
  {Comerford}, {Comparat}, {da Costa}, {Covey}, {Crane}, {Cruz-Gonzalez},
  {Culhane}, {Cunha}, {Dai}, {Damke}, {Darling}, {Davidson}, {Davies},
  {Dawson}, {De Lee}, {Diamond-Stanic}, {Cano-D{\'\i}az}, {S{\'a}nchez},
  {Donor}, {Duckworth}, {Dwelly}, {Eisenstein}, {Elsworth}, {Emsellem},
  {Eracleous}, {Escoffier}, {Fan}, {Farr}, {Feng}, {Fern{\'a}ndez-Trincado},
  {Feuillet}, {Filipp}, {Fillingham}, {Frinchaboy}, {Fromenteau}, {Galbany},
  {Garc{\'\i}a}, {Garc{\'\i}a-Hern{\'a}ndez}, {Ge}, {Geisler}, {Gelfand},
  {G{\'e}ron}, {Gibson}, {Goddy}, {Godoy-Rivera}, {Grabowski}, {Green},
  {Greener}, {Grier}, {Griffith}, {Guo}, {Guy}, {Hadjara}, {Harding},
  {Hasselquist}, {Hayes}, {Hearty}, {Hern{\'a}ndez}, {Hill}, {Hogg},
  {Holtzman}, {Horta}, {Hsieh}, {Hsu}, {Hsu}, {Huber}, {Huertas-Company},
  {Hutchinson}, {Hwang}, {Ibarra-Medel}, {Chitham}, {Ilha}, {Imig}, {Jaekle},
  {Jayasinghe}, {Ji}, {Johnson}, {Jones}, {J{\"o}nsson}, {Katkov}, {Khalatyan},
  {Kinemuchi}, {Kisku}, {Knapen}, {Kneib}, {Kollmeier}, {Kong}, {Kounkel},
  {Kreckel}, {Krishnarao}, {Lacerna}, {Lane}, {Langgin}, {Lavender}, {Law},
  {Lazarz}, {Leung}, {Leung}, {Lewis}, {Li}, {Li}, {Lian}, {Liang}, {Lin},
  {Lin}, {Lin}, {Lintott}, {Long}, {Longa-Pe{\~n}a}, {L{\'o}pez-Cob{\'a}},
  {Lu}, {Lundgren}, {Luo}, {Mackereth}, {de la Macorra}, {Mahadevan},
  {Majewski}, {Manchado}, {Mandeville}, {Maraston}, {Margalef-Bentabol},
  {Masseron}, {Masters}, {Mathur}, {McDermid}, {Mckay}, {Merloni},
  {Merrifield}, {Meszaros}, {Miglio}, {Di Mille}, {Minniti}, {Minsley},
  {Monachesi}, {Moon}, {Mosser}, {Mulchaey}, {Muna}, {Mu{\~n}oz}, {Myers},
  {Myers}, {Nadathur}, {Nair}, {Nandra}, {Neumann}, {Newman}, {Nidever},
  {Nikakhtar}, {Nitschelm}, {O'Connell}, {Garma-Oehmichen}, {Luan Souza de
  Oliveira}, {Olney}, {Oravetz}, {Ortigoza-Urdaneta}, {Osorio}, {Otter},
  {Pace}, {Padilla}, {Pan}, {Pan}, {Parikh}, {Parker}, {Peirani}, {Pe{\~n}a
  Ram{\'\i}rez}, {Penny}, {Percival}, {Perez-Fournon}, {Pinsonneault},
  {Poidevin}, {Poovelil}, {Price-Whelan}, {B{\'a}rbara de Andrade Queiroz},
  {Raddick}, {Ray}, {Rembold}, {Riddle}, {Riffel}, {Riffel}, {Rix}, {Robin},
  {Rodr{\'\i}guez-Puebla}, {Roman-Lopes}, {Rom{\'a}n-Z{\'u}{\~n}iga}, {Rose},
  {Ross}, {Rossi}, {Rubin}, {Salvato}, {S{\'a}nchez}, {S{\'a}nchez-Gallego},
  {Sanderson}, {Santana Rojas}, {Sarceno}, {Sarmiento}, {Sayres}, {Sazonova},
  {Schaefer}, {Schiavon}, {Schlegel}, {Schneider}, {Schultheis}, {Schwope},
  {Serenelli}, {Serna}, {Shao}, {Shapiro}, {Sharma}, {Shen}, {Shetrone}, {Shu},
  {Simon}, {Skrutskie}, {Smethurst}, {Smith}, {Sobeck}, {Spoo}, {Sprague},
  {Stark}, {Stassun}, {Steinmetz}, {Stello}, {Stone-Martinez},
  {Storchi-Bergmann}, {Stringfellow}, {Stutz}, {Su}, {Taghizadeh-Popp},
  {Talbot}, {Tayar}, {Telles}, {Teske}, {Thakar}, {Theissen}, {Tkachenko},
  {Thomas}, {Tojeiro}, {Hernandez Toledo}, {Troup}, {Trump}, {Trussler},
  {Turner}, {Tuttle}, {Unda-Sanzana}, {V{\'a}zquez-Mata}, {Valentini},
  {Valenzuela}, {Vargas-Gonz{\'a}lez}, {Vargas-Maga{\~n}a}, {Alfaro},
  {Villanova}, {Vincenzo}, {Wake}, {Warfield}, {Washington}, {Weaver},
  {Weijmans}, {Weinberg}, {Weiss}, {Westfall}, {Wild}, {Wilde}, {Wilson},
  {Wilson}, {Wilson}, {Wolf}, {Wood-Vasey}, {Yan}, {Zamora}, {Zasowski},
  {Zhang}, {Zhao}, {Zheng}, {Zheng}, \& {Zhu}}]{apogee}
{Abdurro'uf}, {Accetta}, K., {Aerts}, C., {et~al.} 2022, \apjs, 259, 35,
  \dodoi{10.3847/1538-4365/ac4414}

\bibitem[{{Ahumada} {et~al.}(2020){Ahumada}, {Allende Prieto}, {Almeida},
  {Anders}, {Anderson}, {Andrews}, {Anguiano}, {Arcodia}, {Armengaud},
  {Aubert}, {Avila}, {Avila-Reese}, {Badenes}, {Balland}, {Barger},
  {Barrera-Ballesteros}, {Basu}, {Bautista}, {Beaton}, {Beers}, {Benavides},
  {Bender}, {Bernardi}, {Bershady}, {Beutler}, {Bidin}, {Bird}, {Bizyaev},
  {Blanc}, {Blanton}, {Boquien}, {Borissova}, {Bovy}, {Brandt}, {Brinkmann},
  {Brownstein}, {Bundy}, {Bureau}, {Burgasser}, {Burtin}, {Cano-D{\'\i}az},
  {Capasso}, {Cappellari}, {Carrera}, {Chabanier}, {Chaplin}, {Chapman},
  {Cherinka}, {Chiappini}, {Doohyun Choi}, {Chojnowski}, {Chung}, {Clerc},
  {Coffey}, {Comerford}, {Comparat}, {da Costa}, {Cousinou}, {Covey}, {Crane},
  {Cunha}, {Ilha}, {Dai}, {Damsted}, {Darling}, {Davidson}, {Davies}, {Dawson},
  {De}, {de la Macorra}, {De Lee}, {Queiroz}, {Deconto Machado}, {de la Torre},
  {Dell'Agli}, {du Mas des Bourboux}, {Diamond-Stanic}, {Dillon}, {Donor},
  {Drory}, {Duckworth}, {Dwelly}, {Ebelke}, {Eftekharzadeh}, {Davis Eigenbrot},
  {Elsworth}, {Eracleous}, {Erfanianfar}, {Escoffier}, {Fan}, {Farr},
  {Fern{\'a}ndez-Trincado}, {Feuillet}, {Finoguenov}, {Fofie},
  {Fraser-McKelvie}, {Frinchaboy}, {Fromenteau}, {Fu}, {Galbany}, {Garcia},
  {Garc{\'\i}a-Hern{\'a}ndez}, {Garma Oehmichen}, {Ge}, {Geimba Maia},
  {Geisler}, {Gelfand}, {Goddy}, {Gonzalez-Perez}, {Grabowski}, {Green},
  {Grier}, {Guo}, {Guy}, {Harding}, {Hasselquist}, {Hawken}, {Hayes}, {Hearty},
  {Hekker}, {Hogg}, {Holtzman}, {Horta}, {Hou}, {Hsieh}, {Huber}, {Hunt}, {Ider
  Chitham}, {Imig}, {Jaber}, {Jimenez Angel}, {Johnson}, {Jones},
  {J{\"o}nsson}, {Jullo}, {Kim}, {Kinemuchi}, {Kirkpatrick}, {Kite}, {Klaene},
  {Kneib}, {Kollmeier}, {Kong}, {Kounkel}, {Krishnarao}, {Lacerna}, {Lan},
  {Lane}, {Law}, {Le Goff}, {Leung}, {Lewis}, {Li}, {Lian}, {Lin}, {Long},
  {Longa-Pe{\~n}a}, {Lundgren}, {Lyke}, {Mackereth}, {MacLeod}, {Majewski},
  {Manchado}, {Maraston}, {Martini}, {Masseron}, {Masters}, {Mathur},
  {McDermid}, {Merloni}, {Merrifield}, {M{\'e}sz{\'a}ros}, {Miglio}, {Minniti},
  {Minsley}, {Miyaji}, {Mohammad}, {Mosser}, {Mueller}, {Muna},
  {Mu{\~n}oz-Guti{\'e}rrez}, {Myers}, {Nadathur}, {Nair}, {Nandra}, {Correa do
  Nascimento}, {Nevin}, {Newman}, {Nidever}, {Nitschelm}, {Noterdaeme},
  {O'Connell}, {Olmstead}, {Oravetz}, {Oravetz}, {Osorio}, {Pace}, {Padilla},
  {Palanque-Delabrouille}, {Palicio}, {Pan}, {Pan}, {Parker}, {Paviot},
  {Peirani}, {Ram{\'r}ez}, {Penny}, {Percival}, {Perez-Fournon},
  {P{\'e}rez-R{\`a}fols}, {Petitjean}, {Pieri}, {Pinsonneault}, {Poovelil},
  {Povick}, {Prakash}, {Price-Whelan}, {Raddick}, {Raichoor}, {Ray}, {Rembold},
  {Rezaie}, {Riffel}, {Riffel}, {Rix}, {Robin}, {Roman-Lopes},
  {Rom{\'a}n-Z{\'u}{\~n}iga}, {Rose}, {Ross}, {Rossi}, {Rowlands}, {Rubin},
  {Salvato}, {S{\'a}nchez}, {S{\'a}nchez-Menguiano}, {S{\'a}nchez-Gallego},
  {Sayres}, {Schaefer}, {Schiavon}, {Schimoia}, {Schlafly}, {Schlegel},
  {Schneider}, {Schultheis}, {Schwope}, {Seo}, {Serenelli}, {Shafieloo},
  {Shamsi}, {Shao}, {Shen}, {Shetrone}, {Shirley}, {Silva Aguirre}, {Simon},
  {Skrutskie}, {Slosar}, {Smethurst}, {Sobeck}, {Sodi}, {Souto}, {Stark},
  {Stassun}, {Steinmetz}, {Stello}, {Stermer}, {Storchi-Bergmann},
  {Streblyanska}, {Stringfellow}, {Stutz}, {Su{\'a}rez}, {Sun},
  {Taghizadeh-Popp}, {Talbot}, {Tayar}, {Thakar}, {Theriault}, {Thomas},
  {Thomas}, {Tinker}, {Tojeiro}, {Toledo}, {Tremonti}, {Troup}, {Tuttle},
  {Unda-Sanzana}, {Valentini}, {Vargas-Gonz{\'a}lez}, {Vargas-Maga{\~n}a},
  {V{\'a}zquez-Mata}, {Vivek}, {Wake}, {Wang}, {Weaver}, {Weijmans}, {Wild},
  {Wilson}, {Wilson}, {Wolthuis}, {Wood-Vasey}, {Yan}, {Yang}, {Y{\`e}che},
  {Zamora}, {Zarrouk}, {Zasowski}, {Zhang}, {Zhao}, {Zhao}, {Zheng}, {Zheng},
  {Zhu}, \& {Zou}}]{apogeedr16}
{Ahumada}, R., {Allende Prieto}, C., {Almeida}, A., {et~al.} 2020, \apjs, 249,
  3, \dodoi{10.3847/1538-4365/ab929e}

\bibitem[{{Armillotta} {et~al.}(2018){Armillotta}, {Krumholz}, \&
  {Fujimoto}}]{Armillotta2018}
{Armillotta}, L., {Krumholz}, M.~R., \& {Fujimoto}, Y. 2018, \mnras, 481, 5000,
  \dodoi{10.1093/mnras/sty2625}

\bibitem[{{Astropy Collaboration} {et~al.}(2013){Astropy Collaboration},
  {Robitaille}, {Tollerud}, {Greenfield}, {Droettboom}, {Bray}, {Aldcroft},
  {Davis}, {Ginsburg}, {Price-Whelan}, {Kerzendorf}, {Conley}, {Crighton},
  {Barbary}, {Muna}, {Ferguson}, {Grollier}, {Parikh}, {Nair}, {Unther},
  {Deil}, {Woillez}, {Conseil}, {Kramer}, {Turner}, {Singer}, {Fox}, {Weaver},
  {Zabalza}, {Edwards}, {Azalee Bostroem}, {Burke}, {Casey}, {Crawford},
  {Dencheva}, {Ely}, {Jenness}, {Labrie}, {Lim}, {Pierfederici}, {Pontzen},
  {Ptak}, {Refsdal}, {Servillat}, \& {Streicher}}]{Astropy}
{Astropy Collaboration}, {Robitaille}, T.~P., {Tollerud}, E.~J., {et~al.} 2013,
  \aap, 558, A33, \dodoi{10.1051/0004-6361/201322068}

\bibitem[{{Bedell} {et~al.}(2018){Bedell}, {Bean}, {Mel{\'e}ndez}, {Spina},
  {Ram{\'\i}rez}, {Asplund}, {Alves-Brito}, {dos Santos}, {Dreizler}, {Yong},
  {Monroe}, \& {Casagrande}}]{Bedell2018}
{Bedell}, M., {Bean}, J.~L., {Mel{\'e}ndez}, J., {et~al.} 2018, \apj, 865, 68,
  \dodoi{10.3847/1538-4357/aad908}

\bibitem[{{Bell} {et~al.}(1976){Bell}, {Eriksson}, {Gustafsson}, \&
  {Nordlund}}]{Bell1976}
{Bell}, R.~A., {Eriksson}, K., {Gustafsson}, B., \& {Nordlund}, A. 1976, \aaps,
  23, 37

\bibitem[{{Belokurov} {et~al.}(2020){Belokurov}, {Penoyre}, {Oh}, {Iorio},
  {Hodgkin}, {Evans}, {Everall}, {Koposov}, {Tout}, {Izzard}, {Clarke}, \&
  {Brown}}]{ruwe}
{Belokurov}, V., {Penoyre}, Z., {Oh}, S., {et~al.} 2020, \mnras, 496, 1922,
  \dodoi{10.1093/mnras/staa1522}

\bibitem[{{Bland-Hawthorn} {et~al.}(2010){Bland-Hawthorn}, {Krumholz}, \&
  {Freeman}}]{BlandHawthorn2010}
{Bland-Hawthorn}, J., {Krumholz}, M.~R., \& {Freeman}, K. 2010, \apj, 713, 166,
  \dodoi{10.1088/0004-637X/713/1/166}

\bibitem[{{Bovy}(2015)}]{Bovy2015}
{Bovy}, J. 2015, \apjs, 216, 29, \dodoi{10.1088/0067-0049/216/2/29}

\bibitem[{{Bovy}(2016)}]{Bovy2016a}
---. 2016, \apj, 817, 49, \dodoi{10.3847/0004-637X/817/1/49}

\bibitem[{{Bovy} {et~al.}(2016){Bovy}, {Bahmanyar}, {Fritz}, \&
  {Kallivayalil}}]{Bovy2016b}
{Bovy}, J., {Bahmanyar}, A., {Fritz}, T.~K., \& {Kallivayalil}, N. 2016, \apj,
  833, 31, \dodoi{10.3847/1538-4357/833/1/31}

\bibitem[{{Buder} {et~al.}(2018){Buder}, {Asplund}, {Duong}, {Kos}, {Lind},
  {Ness}, {Sharma}, {Bland -Hawthorn}, {Casey}, {de Silva}, {D'Orazi},
  {Freeman}, {Lewis}, {Lin}, {Martell}, {Schlesinger}, {Simpson}, {Zucker},
  {Zwitter}, {Amarsi}, {Anguiano}, {Carollo}, {Casagrande}, {{\v{C}}otar},
  {Cottrell}, {da Costa}, {Gao}, {Hayden}, {Horner}, {Ireland}, {Kafle},
  {Munari}, {Nataf}, {Nordlander}, {Stello}, {Ting}, {Traven}, {Watson},
  {Wittenmyer}, {Wyse}, {Yong}, {Zinn}, {{\v{Z}}erjal}, \& {Galah
  Collaboration}}]{Buder2018}
{Buder}, S., {Asplund}, M., {Duong}, L., {et~al.} 2018, \mnras, 478, 4513,
  \dodoi{10.1093/mnras/sty1281}

\bibitem[{{Buder} {et~al.}(2021){Buder}, {Sharma}, {Kos}, {Amarsi},
  {Nordlander}, {Lind}, {Martell}, {Asplund}, {Bland-Hawthorn}, {Casey}, {de
  Silva}, {D'Orazi}, {Freeman}, {Hayden}, {Lewis}, {Lin}, {Schlesinger},
  {Simpson}, {Stello}, {Zucker}, {Zwitter}, {Beeson}, {Buck}, {Casagrande},
  {Clark}, {{\v{C}}otar}, {da Costa}, {de Grijs}, {Feuillet}, {Horner},
  {Kafle}, {Khanna}, {Kobayashi}, {Liu}, {Montet}, {Nandakumar}, {Nataf},
  {Ness}, {Spina}, {Tepper-Garc{\'\i}a}, {Ting}, {Traven},
  {Vogrin{\v{c}}i{\v{c}}}, {Wittenmyer}, {Wyse}, {{\v{Z}}erjal},
  {{\v{Z}}erjal}, \& {Galah Collaboration}}]{GALAHDR3}
{Buder}, S., {Sharma}, S., {Kos}, J., {et~al.} 2021, \mnras, 506, 150,
  \dodoi{10.1093/mnras/stab1242}

\bibitem[{{Cantat-Gaudin} {et~al.}(2018){Cantat-Gaudin}, {Jordi}, {Vallenari},
  {Bragaglia}, {Balaguer-N{\'u}{\~n}ez}, {Soubiran}, {Bossini}, {Moitinho},
  {Castro-Ginard}, {Krone-Martins}, {Casamiquela}, {Sordo}, \&
  {Carrera}}]{Cantat-Gaudin2018}
{Cantat-Gaudin}, T., {Jordi}, C., {Vallenari}, A., {et~al.} 2018, \aap, 618,
  A93, \dodoi{10.1051/0004-6361/201833476}

\bibitem[{{Casali} {et~al.}(2020){Casali}, {Spina}, {Magrini}, {Karakas},
  {Kobayashi}, {Casey}, {Feltzing}, {Van der Swaelmen}, {Tsantaki},
  {Jofr{\'e}}, {Bragaglia}, {Feuillet}, {Bensby}, {Biazzo}, {Gonneau},
  {Tautvai{\v{s}}ien{\.{e}}}, {Baratella}, {Roccatagliata}, {Pancino}, {Sousa},
  {Adibekyan}, {Martell}, {Bayo}, {Jackson}, {Jeffries}, {Gilmore}, {Randich},
  {Alfaro}, {Koposov}, {Korn}, {Recio-Blanco}, {Smiljanic}, {Franciosini},
  {Hourihane}, {Monaco}, {Morbidelli}, {Sacco}, {Worley}, \&
  {Zaggia}}]{Casali2020}
{Casali}, G., {Spina}, L., {Magrini}, L., {et~al.} 2020, \aap, 639, A127,
  \dodoi{10.1051/0004-6361/202038055}

\bibitem[{{Cheng} {et~al.}(2021){Cheng}, {Price-Jones}, \& {Bovy}}]{Cheng2020}
{Cheng}, C.~M., {Price-Jones}, N., \& {Bovy}, J. 2021, \mnras, 506, 5573,
  \dodoi{10.1093/mnras/stab2106}

\bibitem[{{Conroy} {et~al.}(2019){Conroy}, {Naidu}, {Zaritsky}, {Bonaca},
  {Cargile}, {Johnson}, \& {Caldwell}}]{H3}
{Conroy}, C., {Naidu}, R.~P., {Zaritsky}, D., {et~al.} 2019, \apj, 887, 237,
  \dodoi{10.3847/1538-4357/ab5710}

\bibitem[{{Cox} {et~al.}(2012){Cox}, {Kerschbaum}, {van Marle}, {Decin},
  {Ladjal}, {Mayer}, {Groenewegen}, {van Eck}, {Royer}, {Ottensamer}, {Ueta},
  {Jorissen}, {Mecina}, {Meliani}, {Luntzer}, {Blommaert}, {Posch},
  {Vandenbussche}, \& {Waelkens}}]{Cox2012}
{Cox}, N.~L.~J., {Kerschbaum}, F., {van Marle}, A.~J., {et~al.} 2012, \aap,
  537, A35, \dodoi{10.1051/0004-6361/201117910}

\bibitem[{{Cui} {et~al.}(2012){Cui}, {Zhao}, {Chu}, {Li}, {Li}, {Zhang}, {Su},
  {Yao}, {Wang}, {Xing}, {Li}, {Zhu}, {Wang}, {Gu}, {Luo}, {Xu}, {Zhang},
  {Liu}, {Zhang}, {Yang}, {Cao}, {Chen}, {Chen}, {Chen}, {Chen}, {Chu}, {Feng},
  {Gong}, {Hou}, {Hu}, {Hu}, {Hu}, {Jia}, {Jiang}, {Jiang}, {Jiang}, {Jin},
  {Li}, {Li}, {Li}, {Liu}, {Liu}, {Lu}, {Mao}, {Men}, {Qi}, {Qi}, {Shi},
  {Tang}, {Tao}, {Wang}, {Wang}, {Wang}, {Wang}, {Wang}, {Wang}, {Wang},
  {Wang}, {Wang}, {Wang}, {Wang}, {Wang}, {Xu}, {Xu}, {Yang}, {Yu}, {Yuan},
  {Yuan}, {Zhai}, {Zhang}, {Zhang}, {Zhang}, {Zhao}, {Zhou}, {Zhou}, {Zhu}, \&
  {Zou}}]{LAMOST}
{Cui}, X.-Q., {Zhao}, Y.-H., {Chu}, Y.-Q., {et~al.} 2012, Research in Astronomy
  and Astrophysics, 12, 1197, \dodoi{10.1088/1674-4527/12/9/003}

\bibitem[{{de Mijolla} {et~al.}(2021){de Mijolla}, {Ness}, {Viti}, \&
  {Wheeler}}]{deMijolla2021}
{de Mijolla}, D., {Ness}, M., {Viti}, S., \& {Wheeler}, A. 2021, arXiv
  e-prints, arXiv:2103.06377.
\newblock \doarXiv{2103.06377}

\bibitem[{{De Silva} {et~al.}(2015){De Silva}, {Freeman}, {Bland-Hawthorn},
  {Martell}, {de Boer}, {Asplund}, {Keller}, {Sharma}, {Zucker}, {Zwitter},
  {Anguiano}, {Bacigalupo}, {Bayliss}, {Beavis}, {Bergemann}, {Campbell},
  {Cannon}, {Carollo}, {Casagrande}, {Casey}, {Da Costa}, {D'Orazi}, {Dotter},
  {Duong}, {Heger}, {Ireland}, {Kafle}, {Kos}, {Lattanzio}, {Lewis}, {Lin},
  {Lind}, {Munari}, {Nataf}, {O'Toole}, {Parker}, {Reid}, {Schlesinger},
  {Sheinis}, {Simpson}, {Stello}, {Ting}, {Traven}, {Watson}, {Wittenmyer},
  {Yong}, \& {{\v{Z}}erjal}}]{DeSilva2015}
{De Silva}, G.~M., {Freeman}, K.~C., {Bland-Hawthorn}, J., {et~al.} 2015,
  \mnras, 449, 2604, \dodoi{10.1093/mnras/stv327}

\bibitem[{{Emerick} {et~al.}(2020){Emerick}, {Bryan}, \& {Mac
  Low}}]{Emerick2020}
{Emerick}, A., {Bryan}, G.~L., \& {Mac Low}, M.-M. 2020, arXiv e-prints,
  arXiv:2007.03702, \dodoi{10.48550/arXiv.2007.03702}

\bibitem[{{Feltzing} {et~al.}(2017){Feltzing}, {Howes}, {McMillan}, \&
  {Stonkut{\.{e}}}}]{Feltzing2017}
{Feltzing}, S., {Howes}, L.~M., {McMillan}, P.~J., \& {Stonkut{\.{e}}}, E.
  2017, \mnras, 465, L109, \dodoi{10.1093/mnrasl/slw209}

\bibitem[{{Fischer} \& {Valenti}(2005)}]{fischer05}
{Fischer}, D.~A., \& {Valenti}, J. 2005, \apj, 622, 1102,
  \dodoi{10.1086/428383}

\bibitem[{{Freeman} \& {Bland-Hawthorn}(2002)}]{Freeman2002}
{Freeman}, K., \& {Bland-Hawthorn}, J. 2002, \araa, 40, 487,
  \dodoi{10.1146/annurev.astro.40.060401.093840}

\bibitem[{{Gaia Collaboration} {et~al.}(2020){Gaia Collaboration}, {Brown},
  {Vallenari}, {Prusti}, {de Bruijne}, {Babusiaux}, \& {Biermann}}]{GaiaEDR3}
{Gaia Collaboration}, {Brown}, A.~G.~A., {Vallenari}, A., {et~al.} 2020, arXiv
  e-prints, arXiv:2012.01533.
\newblock \doarXiv{2012.01533}

\bibitem[{{Gilmore} {et~al.}(2022){Gilmore}, {Randich}, {Worley}, {Hourihane},
  {Gonneau}, {Sacco}, {Lewis}, {Magrini}, {Fran{\c{c}}ois}, {Jeffries},
  {Koposov}, {Bragaglia}, {Alfaro}, {Allende Prieto}, {Blomme}, {Korn},
  {Lanzafame}, {Pancino}, {Recio-Blanco}, {Smiljanic}, {Van Eck}, {Zwitter},
  {Bensby}, {Flaccomio}, {Irwin}, {Franciosini}, {Morbidelli}, {Damiani},
  {Bonito}, {Friel}, {Vink}, {Prisinzano}, {Abbas}, {Hatzidimitriou}, {Held},
  {Jordi}, {Paunzen}, {Spagna}, {Jackson}, {Ma{\'\i}z Apell{\'a}niz},
  {Asplund}, {Bonifacio}, {Feltzing}, {Binney}, {Drew}, {Ferguson}, {Micela},
  {Negueruela}, {Prusti}, {Rix}, {Vallenari}, {Bergemann}, {Casey}, {de
  Laverny}, {Frasca}, {Hill}, {Lind}, {Sbordone}, {Sousa}, {Adibekyan},
  {Caffau}, {Daflon}, {Feuillet}, {Gebran}, {Gonzalez Hernandez}, {Guiglion},
  {Herrero}, {Lobel}, {Merle}, {Mikolaitis}, {Montes}, {Morel}, {Ruchti},
  {Soubiran}, {Tabernero}, {Tautvai{\v{s}}ien{\.{e}}}, {Traven}, {Valentini},
  {Van der Swaelmen}, {Villanova}, {Viscasillas V{\'a}zquez}, {Bayo}, {Biazzo},
  {Carraro}, {Edvardsson}, {Heiter}, {Jofr{\'e}}, {Marconi}, {Martayan},
  {Masseron}, {Monaco}, {Walton}, {Zaggia}, {Aguirre B{\o}rsen-Koch}, {Alves},
  {Balaguer-Nunez}, {Barklem}, {Barrado}, {Bellazzini}, {Berlanas}, {Binks},
  {Bressan}, {Capuzzo-Dolcetta}, {Casagrande}, {Casamiquela}, {Collins},
  {D'Orazi}, {Dantas}, {Debattista}, {Delgado-Mena}, {Di Marcantonio},
  {Drazdauskas}, {Evans}, {Famaey}, {Franchini}, {Fr{\'e}mat}, {Fu}, {Geisler},
  {Gerhard}, {Gonz{\'a}lez Solares}, {Grebel}, {Guti{\'e}rrez Albarr{\'a}n},
  {Jim{\'e}nez-Esteban}, {J{\"o}nsson}, {Khachaturyants}, {Kordopatis}, {Kos},
  {Lagarde}, {Ludwig}, {Mahy}, {Mapelli}, {Marfil}, {Martell}, {Messina},
  {Miglio}, {Minchev}, {Moitinho}, {Montalban}, {Monteiro}, {Morossi},
  {Mowlavi}, {Mucciarelli}, {Murphy}, {Nardetto}, {Ortolani}, {Paletou},
  {Palou{\v{s}}}, {Pickering}, {Quirrenbach}, {Re Fiorentin}, {Read}, {Romano},
  {Ryde}, {Sanna}, {Santos}, {Seabroke}, {Spina}, {Steinmetz}, {Stonkut{\'e}},
  {Sutorius}, {Th{\'e}venin}, {Tosi}, {Tsantaki}, {Wright}, {Wyse}, {Zoccali},
  {Zorec}, \& {Zucker}}]{Gaia-ESO}
{Gilmore}, G., {Randich}, S., {Worley}, C.~C., {et~al.} 2022, \aap, 666, A120,
  \dodoi{10.1051/0004-6361/202243134}

\bibitem[{{Griffith} {et~al.}(2023){Griffith}, {Hogg}, {Dalcanton},
  {Hasselquist}, {Ratcilffe}, {Ness}, \& {Weinberg}}]{Griffith2023}
{Griffith}, E.~J., {Hogg}, D.~W., {Dalcanton}, J.~J., {et~al.} 2023, arXiv
  e-prints, arXiv:2307.05691, \dodoi{10.48550/arXiv.2307.05691}

\bibitem[{{Griffith} {et~al.}(2022){Griffith}, {Weinberg}, {Buder}, {Johnson},
  {Johnson}, \& {Vincenzo}}]{Griffith2022}
{Griffith}, E.~J., {Weinberg}, D.~H., {Buder}, S., {et~al.} 2022, \apj, 931,
  23, \dodoi{10.3847/1538-4357/ac5826}

\bibitem[{{Gustafsson} {et~al.}(1975){Gustafsson}, {Bell}, {Eriksson}, \&
  {Nordlund}}]{Gustafsson1975}
{Gustafsson}, B., {Bell}, R.~A., {Eriksson}, K., \& {Nordlund}, A. 1975, \aap,
  500, 67

\bibitem[{{Gustafsson} {et~al.}(2008){Gustafsson}, {Edvardsson}, {Eriksson},
  {J{\o}rgensen}, {Nordlund}, \& {Plez}}]{Gustafsson2008}
{Gustafsson}, B., {Edvardsson}, B., {Eriksson}, K., {et~al.} 2008, \aap, 486,
  951, \dodoi{10.1051/0004-6361:200809724}

\bibitem[{{Halevi} \& {M{\"o}sta}(2018)}]{Halevi2018}
{Halevi}, G., \& {M{\"o}sta}, P. 2018, \mnras, 477, 2366,
  \dodoi{10.1093/mnras/sty797}

\bibitem[{{Hawkins} {et~al.}(2020){Hawkins}, {Lucey}, {Ting}, {Ji}, {Katzberg},
  {Thompson}, {El-Badry}, {Teske}, {Nelson}, \& {Carrillo}}]{Hawkins2020}
{Hawkins}, K., {Lucey}, M., {Ting}, Y.-S., {et~al.} 2020, \mnras, 492, 1164,
  \dodoi{10.1093/mnras/stz3132}

\bibitem[{{Ho} {et~al.}(2017){Ho}, {Ness}, {Hogg}, {Rix}, {Liu}, {Yang},
  {Zhang}, {Hou}, \& {Wang}}]{Cannon}
{Ho}, A. Y.~Q., {Ness}, M.~K., {Hogg}, D.~W., {et~al.} 2017, \apj, 836, 5,
  \dodoi{10.3847/1538-4357/836/1/5}

\bibitem[{{Holtzman} {et~al.}(2018){Holtzman}, {Hasselquist}, {Shetrone},
  {Cunha}, {Allende Prieto}, {Anguiano}, {Bizyaev}, {Bovy}, {Casey},
  {Edvardsson}, {Johnson}, {J{\"o}nsson}, {Meszaros}, {Smith}, {Sobeck},
  {Zamora}, {Chojnowski}, {Fernandez-Trincado}, {Garcia-Hernandez}, {Majewski},
  {Pinsonneault}, {Souto}, {Stringfellow}, {Tayar}, {Troup}, \&
  {Zasowski}}]{apogeedr13}
{Holtzman}, J.~A., {Hasselquist}, S., {Shetrone}, M., {et~al.} 2018, \aj, 156,
  125, \dodoi{10.3847/1538-3881/aad4f9}

\bibitem[{{Hunter}(2007)}]{matplotlib}
{Hunter}, J.~D. 2007, Computing in Science and Engineering, 9, 90,
  \dodoi{10.1109/MCSE.2007.55}

\bibitem[{{Jofr{\'e}} {et~al.}(2019){Jofr{\'e}}, {Heiter}, \&
  {Soubiran}}]{Jofre2019}
{Jofr{\'e}}, P., {Heiter}, U., \& {Soubiran}, C. 2019, \araa, 57, 571,
  \dodoi{10.1146/annurev-astro-091918-104509}

\bibitem[{{Jofr{\'e}} {et~al.}(2017){Jofr{\'e}}, {Heiter}, {Worley},
  {Blanco-Cuaresma}, {Soubiran}, {Masseron}, {Hawkins}, {Adibekyan}, {Buder},
  {Casamiquela}, {Gilmore}, {Hourihane}, \& {Tabernero}}]{Jofre2017}
{Jofr{\'e}}, P., {Heiter}, U., {Worley}, C.~C., {et~al.} 2017, \aap, 601, A38,
  \dodoi{10.1051/0004-6361/201629833}

\bibitem[{{Karakas} \& {Lattanzio}(2014)}]{Karakas2014}
{Karakas}, A.~I., \& {Lattanzio}, J.~C. 2014, \pasa, 31, e030,
  \dodoi{10.1017/pasa.2014.21}

\bibitem[{{Kobayashi} {et~al.}(2020){Kobayashi}, {Karakas}, \&
  {Lugaro}}]{Kobayashi2020}
{Kobayashi}, C., {Karakas}, A.~I., \& {Lugaro}, M. 2020, \apj, 900, 179,
  \dodoi{10.3847/1538-4357/abae65}

\bibitem[{{Kobayashi} {et~al.}(2023){Kobayashi}, {Mandel}, {Belczynski},
  {Goriely}, {Janka}, {Just}, {Ruiter}, {Vanbeveren}, {Kruckow}, {Briel},
  {Eldridge}, \& {Stanway}}]{Kobayashi2023}
{Kobayashi}, C., {Mandel}, I., {Belczynski}, K., {et~al.} 2023, \apjl, 943,
  L12, \dodoi{10.3847/2041-8213/acad82}

\bibitem[{{Korobkin} {et~al.}(2012){Korobkin}, {Rosswog}, {Arcones}, \&
  {Winteler}}]{Korobkin2012}
{Korobkin}, O., {Rosswog}, S., {Arcones}, A., \& {Winteler}, C. 2012, \mnras,
  426, 1940, \dodoi{10.1111/j.1365-2966.2012.21859.x}

\bibitem[{{Krumholz} \& {Ting}(2018)}]{Krumholz2018}
{Krumholz}, M.~R., \& {Ting}, Y.-S. 2018, \mnras, 475, 2236,
  \dodoi{10.1093/mnras/stx3286}

\bibitem[{{Lambert} \& {Reddy}(2016)}]{Lambert2016}
{Lambert}, D.~L., \& {Reddy}, A. B.~S. 2016, \apj, 831, 202,
  \dodoi{10.3847/0004-637X/831/2/202}

\bibitem[{{Lian} {et~al.}(2023){Lian}, {Storm}, {Guiglion}, {Serenelli},
  {Cote}, {Karakas}, {Boardman}, \& {Bergemann}}]{Lian2023}
{Lian}, J., {Storm}, N., {Guiglion}, G., {et~al.} 2023, \mnras, 525, 1329,
  \dodoi{10.1093/mnras/stad2390}

\bibitem[{{Marigo} {et~al.}(2017){Marigo}, {Girardi}, {Bressan}, {Rosenfield},
  {Aringer}, {Chen}, {Dussin}, {Nanni}, {Pastorelli}, {Rodrigues}, {Trabucchi},
  {Bladh}, {Dalcanton}, {Groenewegen}, {Montalb{\'a}n}, \& {Wood}}]{Marigo2017}
{Marigo}, P., {Girardi}, L., {Bressan}, A., {et~al.} 2017, \apj, 835, 77,
  \dodoi{10.3847/1538-4357/835/1/77}

\bibitem[{{Nandakumar} {et~al.}(2022){Nandakumar}, {Hayden}, {Sharma}, {Buder},
  {Asplund}, {Bland-Hawthorn}, {De Silva}, {D'Orazi}, {Freeman}, {Kos},
  {Lewis}, {Martell}, {Schlesinger}, {Lin}, {Simpson}, {Zucker}, {Zwitter},
  {Nordlander}, {Casagrande}, {Lind}, {C{\^o}tar}, {Stello}, {Wittenmyer}, \&
  {Tepper-Garcia}}]{Nandakumar2022}
{Nandakumar}, G., {Hayden}, M.~R., {Sharma}, S., {et~al.} 2022, \mnras, 513,
  232, \dodoi{10.1093/mnras/stac873}

\bibitem[{{Nelson} {et~al.}(2021){Nelson}, {Ting}, {Hawkins}, {Ji}, {Kamdar},
  \& {El-Badry}}]{Nelson2021}
{Nelson}, T., {Ting}, Y.-S., {Hawkins}, K., {et~al.} 2021, \apj, 921, 118,
  \dodoi{10.3847/1538-4357/ac14be}

\bibitem[{{Ness} {et~al.}(2015){Ness}, {Hogg}, {Rix}, {Ho}, \&
  {Zasowski}}]{Ness2015}
{Ness}, M., {Hogg}, D.~W., {Rix}, H.~W., {Ho}, A. Y.~Q., \& {Zasowski}, G.
  2015, \apj, 808, 16, \dodoi{10.1088/0004-637X/808/1/16}

\bibitem[{{Ness} {et~al.}(2018){Ness}, {Rix}, {Hogg}, {Casey}, {Holtzman},
  {Fouesneau}, {Zasowski}, {Geisler}, {Shetrone}, {Minniti}, {Frinchaboy}, \&
  {Roman-Lopes}}]{Ness2018}
{Ness}, M., {Rix}, H.~W., {Hogg}, D.~W., {et~al.} 2018, \apj, 853, 198,
  \dodoi{10.3847/1538-4357/aa9d8e}

\bibitem[{{Ness} {et~al.}(2019){Ness}, {Johnston}, {Blancato}, {Rix}, {Beane},
  {Bird}, \& {Hawkins}}]{Ness2019}
{Ness}, M.~K., {Johnston}, K.~V., {Blancato}, K., {et~al.} 2019, \apj, 883,
  177, \dodoi{10.3847/1538-4357/ab3e3c}

\bibitem[{{Ness} {et~al.}(2022){Ness}, {Wheeler}, {McKinnon}, {Horta}, {Casey},
  {Cunningham}, \& {Price-Whelan}}]{Ness2022}
{Ness}, M.~K., {Wheeler}, A.~J., {McKinnon}, K., {et~al.} 2022, \apj, 926, 144,
  \dodoi{10.3847/1538-4357/ac4754}

\bibitem[{{Nielsen} {et~al.}(2023){Nielsen}, {Gent}, {Bergemann}, {Eitner}, \&
  {Johansen}}]{Nielsen2023}
{Nielsen}, J., {Gent}, M.~R., {Bergemann}, M., {Eitner}, P., \& {Johansen}, A.
  2023, arXiv e-prints, arXiv:2308.15504, \dodoi{10.48550/arXiv.2308.15504}

\bibitem[{{Piskunov} \& {Valenti}(2017)}]{Piskunov2017}
{Piskunov}, N., \& {Valenti}, J.~A. 2017, \aap, 597, A16,
  \dodoi{10.1051/0004-6361/201629124}

\bibitem[{{Poovelil} {et~al.}(2020){Poovelil}, {Zasowski}, {Hasselquist},
  {Seth}, {Donor}, {Beaton}, {Cunha}, {Frinchaboy},
  {Garc{\'\i}a-Hern{\'a}ndez}, {Hawkins}, {Kratter}, {Lane}, \&
  {Nitschelm}}]{Poovelil2020}
{Poovelil}, V.~J., {Zasowski}, G., {Hasselquist}, S., {et~al.} 2020, \apj, 903,
  55, \dodoi{10.3847/1538-4357/abb93e}

\bibitem[{{Price-Jones} {et~al.}(2020){Price-Jones}, {Bovy}, {Webb}, {Allende
  Prieto}, {Beaton}, {Brownstein}, {Cohen}, {Cunha}, {Donor}, {Frinchaboy},
  {Garc{\'\i}a-Hern{\'a}ndez}, {Lane}, {Majewski}, {Nidever}, \&
  {Roman-Lopes}}]{Price-Jones2020}
{Price-Jones}, N., {Bovy}, J., {Webb}, J.~J., {et~al.} 2020, \mnras, 496, 5101,
  \dodoi{10.1093/mnras/staa1905}

\bibitem[{{Ratcliffe} {et~al.}(2023){Ratcliffe}, {Minchev}, {Cescutti},
  {Spitoni}, {J{\"o}nsson}, {Anders}, {Queiroz}, \&
  {Steinmetz}}]{Ratcliffe2023}
{Ratcliffe}, B., {Minchev}, I., {Cescutti}, G., {et~al.} 2023, arXiv e-prints,
  arXiv:2307.11159, \dodoi{10.48550/arXiv.2307.11159}

\bibitem[{{Reback} {et~al.}(2020){Reback}, {McKinney}, {Jbrockmendel}, {Den Van
  Bossche}, {Augspurger}, {Cloud}, {Gfyoung}, {Sinhrks}, {Klein}, {Roeschke},
  {Hawkins}, {Tratner}, {She}, {Ayd}, {Petersen}, {Garcia}, {Schendel},
  {Hayden}, {MomIsBestFriend}, {Jancauskas}, {Battiston}, {Seabold},
  {Chris-B1}, {H-Vetinari}, {Hoyer}, {Overmeire}, {Alimcmaster1}, {Dong},
  {Whelan}, \& {Mehyar}}]{pandas}
{Reback}, J., {McKinney}, W., {Jbrockmendel}, {et~al.} 2020,
  {pandas-dev/pandas: Pandas 1.0.3}, v1.0.3,  Zenodo,
  \dodoi{10.5281/zenodo.3509134}

\bibitem[{{Sharma} {et~al.}(2018){Sharma}, {Stello}, {Buder}, {Kos},
  {Bland-Hawthorn}, {Asplund}, {Duong}, {Lin}, {Lind}, {Ness}, {Huber},
  {Zwitter}, {Traven}, {Hon}, {Kafle}, {Khanna}, {Saddon}, {Anguiano}, {Casey},
  {Freeman}, {Martell}, {De Silva}, {Simpson}, {Wittenmyer}, \&
  {Zucker}}]{Sharma2018}
{Sharma}, S., {Stello}, D., {Buder}, S., {et~al.} 2018, \mnras, 473, 2004,
  \dodoi{10.1093/mnras/stx2582}

\bibitem[{{Sharma} {et~al.}(2022){Sharma}, {Hayden}, {Bland-Hawthorn},
  {Stello}, {Buder}, {Zinn}, {Spina}, {Kallinger}, {Asplund}, {De Silva},
  {D'Orazi}, {Freeman}, {Kos}, {Lewis}, {Lin}, {Lind}, {Martell},
  {Schlesinger}, {Simpson}, {Zucker}, {Zwitter}, {Chen}, {Cotar}, {Kafle},
  {Khanna}, {Tepper-Garcia}, {Wang}, \& {Wittenmyer}}]{Sharma2022}
{Sharma}, S., {Hayden}, M.~R., {Bland-Hawthorn}, J., {et~al.} 2022, \mnras,
  510, 734, \dodoi{10.1093/mnras/stab3341}

\bibitem[{{Siegel} {et~al.}(2019){Siegel}, {Barnes}, \& {Metzger}}]{Siegel2019}
{Siegel}, D.~M., {Barnes}, J., \& {Metzger}, B.~D. 2019, \nat, 569, 241,
  \dodoi{10.1038/s41586-019-1136-0}

\bibitem[{{Siegel} \& {Metzger}(2017)}]{Siegel2017}
{Siegel}, D.~M., \& {Metzger}, B.~D. 2017, \prl, 119, 231102,
  \dodoi{10.1103/PhysRevLett.119.231102}

\bibitem[{{Sk{\'u}lad{\'o}ttir} {et~al.}(2019){Sk{\'u}lad{\'o}ttir}, {Hansen},
  {Salvadori}, \& {Choplin}}]{Skuladottier2019}
{Sk{\'u}lad{\'o}ttir}, {\'A}., {Hansen}, C.~J., {Salvadori}, S., \& {Choplin},
  A. 2019, \aap, 631, A171, \dodoi{10.1051/0004-6361/201936125}

\bibitem[{{Spina} {et~al.}(2021){Spina}, {Ting}, {De Silva}, {Frankel},
  {Sharma}, {Cantat-Gaudin}, {Joyce}, {Stello}, {Karakas}, {Asplund},
  {Nordlander}, {Casagrande}, {D'Orazi}, {Casey}, {Cottrell},
  {Tepper-Garc{\'\i}a}, {Baratella}, {Kos}, {{\v{C}}otar}, {Bland-Hawthorn},
  {Buder}, {Freeman}, {Hayden}, {Lewis}, {Lin}, {Lind}, {Martell},
  {Schlesinger}, {Simpson}, {Zucker}, \& {Zwitter}}]{Spina2021}
{Spina}, L., {Ting}, Y.~S., {De Silva}, G.~M., {et~al.} 2021, \mnras, 503,
  3279, \dodoi{10.1093/mnras/stab471}

\bibitem[{{Steinmetz} {et~al.}(2020){Steinmetz}, {Matijevi{\v{c}}}, {Enke},
  {Zwitter}, {Guiglion}, {McMillan}, {Kordopatis}, {Valentini}, {Chiappini},
  {Casagrande}, {Wojno}, {Anguiano}, {Bienaym{\'e}}, {Bijaoui}, {Binney},
  {Burton}, {Cass}, {de Laverny}, {Fiegert}, {Freeman}, {Fulbright}, {Gibson},
  {Gilmore}, {Grebel}, {Helmi}, {Kunder}, {Munari}, {Navarro}, {Parker},
  {Ruchti}, {Recio-Blanco}, {Reid}, {Seabroke}, {Siviero}, {Siebert}, {Stupar},
  {Watson}, {Williams}, {Wyse}, {Anders}, {Antoja}, {Birko}, {Bland-Hawthorn},
  {Bossini}, {Garc{\'\i}a}, {Carrillo}, {Chaplin}, {Elsworth}, {Famaey},
  {Gerhard}, {Jofre}, {Just}, {Mathur}, {Miglio}, {Minchev}, {Monari},
  {Mosser}, {Ritter}, {Rodrigues}, {Scholz}, {Sharma}, {Sysoliatina}, \& {RAVE
  Collaboration}}]{RAVE}
{Steinmetz}, M., {Matijevi{\v{c}}}, G., {Enke}, H., {et~al.} 2020, \aj, 160,
  82, \dodoi{10.3847/1538-3881/ab9ab9}

\bibitem[{{Ting} {et~al.}(2018){Ting}, {Conroy}, {Rix}, \&
  {Asplund}}]{Ting2018}
{Ting}, Y.-S., {Conroy}, C., {Rix}, H.-W., \& {Asplund}, M. 2018, \apj, 860,
  159, \dodoi{10.3847/1538-4357/aac6c9}

\bibitem[{{Ting} {et~al.}(2012){Ting}, {Freeman}, {Kobayashi}, {De Silva}, \&
  {Bland-Hawthorn}}]{Ting2012}
{Ting}, Y.-S., {Freeman}, K.~C., {Kobayashi}, C., {De Silva}, G.~M., \&
  {Bland-Hawthorn}, J. 2012, \mnras, 421, 1231,
  \dodoi{10.1111/j.1365-2966.2011.20387.x}

\bibitem[{{Ting} \& {Weinberg}(2022)}]{Ting2022}
{Ting}, Y.-S., \& {Weinberg}, D.~H. 2022, \apj, 927, 209,
  \dodoi{10.3847/1538-4357/ac5023}

\bibitem[{{Valenti} \& {Piskunov}(1996)}]{Valenti1996}
{Valenti}, J.~A., \& {Piskunov}, N. 1996, \aaps, 118, 595

\bibitem[{{van der Walt} {et~al.}(2011){van der Walt}, {Colbert}, \&
  {Varoquaux}}]{numpy}
{van der Walt}, S., {Colbert}, S.~C., \& {Varoquaux}, G. 2011, Computing in
  Science and Engineering, 13, 22, \dodoi{10.1109/MCSE.2011.37}

\bibitem[{{Virtanen} {et~al.}(2020){Virtanen}, {Gommers}, {Oliphant},
  {Haberland}, {Reddy}, {Cournapeau}, {Burovski}, {Peterson}, {Weckesser},
  {Bright}, {van der Walt}, {Brett}, {Wilson}, {Millman}, {Mayorov}, {Nelson},
  {Jones}, {Kern}, {Larson}, {Carey}, {Polat}, {Feng}, {Moore}, {VanderPlas},
  {Laxalde}, {Perktold}, {Cimrman}, {Henriksen}, {Quintero}, {Harris},
  {Archibald}, {Ribeiro}, {Pedregosa}, {van Mulbregt}, \& {SciPy 1. 0
  Contributors}}]{scipy}
{Virtanen}, P., {Gommers}, R., {Oliphant}, T.~E., {et~al.} 2020, Nature
  Methods, 17, 261, \dodoi{10.1038/s41592-019-0686-2}

\bibitem[{{Vogrin{\v{c}}i{\v{c}}} {et~al.}(2023){Vogrin{\v{c}}i{\v{c}}}, {Kos},
  {Zwitter}, {Traven}, {Beeson}, {{\v{C}}otar}, {Munari}, {Buder}, {Martell},
  {Lewis}, {De Silva}, {Hayden}, {Bland-Hawthorn}, \& {D'Orazi}}]{Vogrin2023}
{Vogrin{\v{c}}i{\v{c}}}, R., {Kos}, J., {Zwitter}, T., {et~al.} 2023, \mnras,
  521, 3727, \dodoi{10.1093/mnras/stad678}

\bibitem[{{Weinberg} {et~al.}(2022){Weinberg}, {Holtzman}, {Johnson}, {Hayes},
  {Hasselquist}, {Shetrone}, {Ting}, {Beaton}, {Beers}, {Bird}, {Bizyaev},
  {Blanton}, {Cunha}, {Fern{\'a}ndez-Trincado}, {Frinchaboy},
  {Garc{\'\i}a-Hern{\'a}ndez}, {Griffith}, {Johnson}, {J{\"o}nsson}, {Lane},
  {Leung}, {Mackereth}, {Majewski}, {M{\'e}sz{\'a}ros}, {Nitschelm}, {Pan},
  {Schiavon}, {Schneider}, {Schultheis}, {Smith}, {Sobeck}, {Stassun},
  {Stringfellow}, {Vincenzo}, {Wilson}, \& {Zasowski}}]{Weinberg2022}
{Weinberg}, D.~H., {Holtzman}, J.~A., {Johnson}, J.~A., {et~al.} 2022, \apjs,
  260, 32, \dodoi{10.3847/1538-4365/ac6028}

\bibitem[{{Wheeler} {et~al.}(2020){Wheeler}, {Ness}, {Buder}, {Bland-Hawthorn},
  {Silva}, {Hayden}, {Kos}, {Lewis}, {Martell}, {Sharma}, {Simpson}, {Zucker},
  \& {Zwitter}}]{Wheeler2020}
{Wheeler}, A., {Ness}, M., {Buder}, S., {et~al.} 2020, \apj, 898, 58,
  \dodoi{10.3847/1538-4357/ab9a46}

\end{thebibliography}
\bibliographystyle{aasjournal}

\begin{figure*}
    \centering
    \includegraphics[angle=90,origin=c, width=14cm]{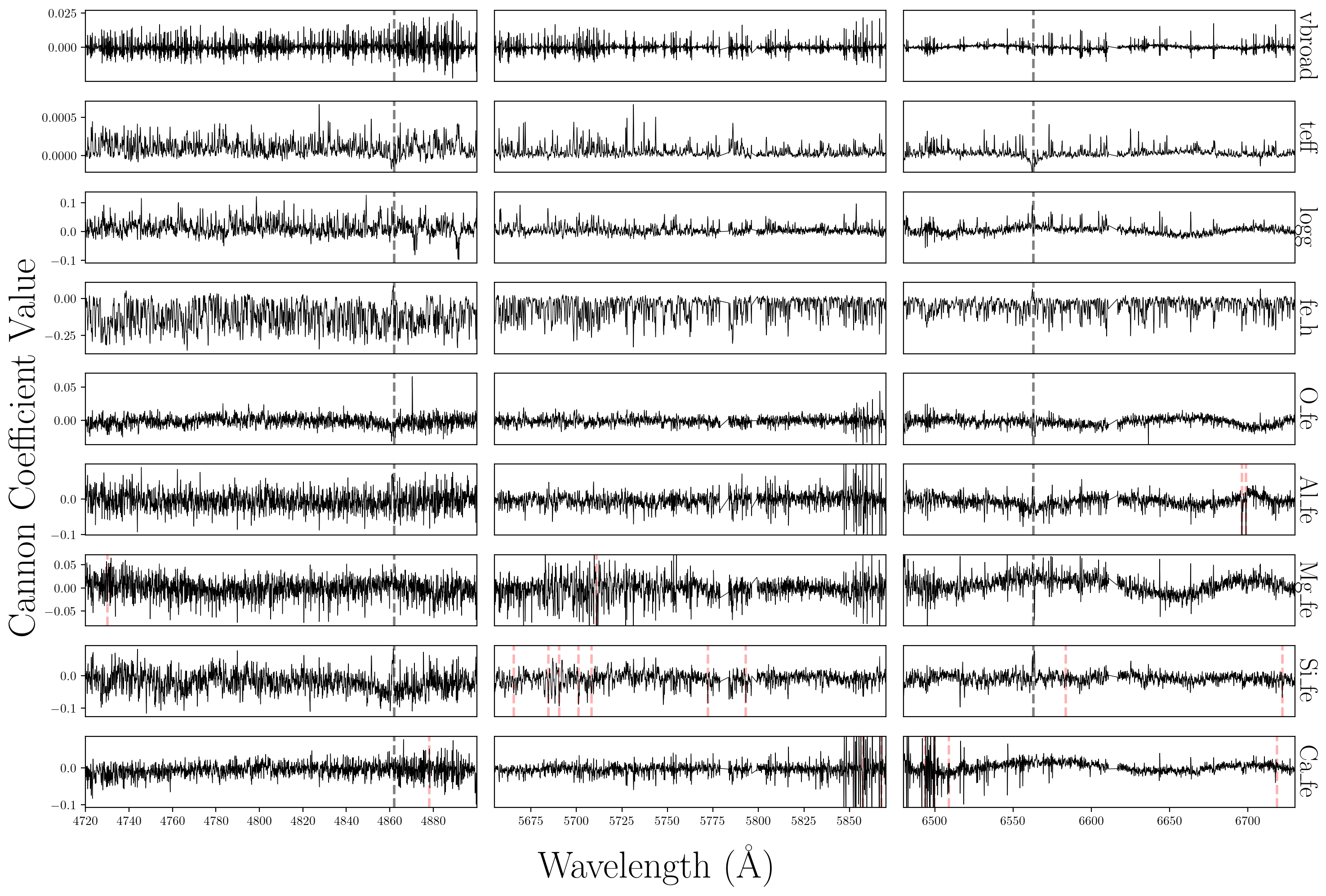}
    \caption{First order \textit{Cannon} model coefficients for the first nine labels as a function of wavelength. Dashed gray lines mark H$\beta$ and H$\alpha$.  Dashed red lines mark known strong lines of the relevant element corresponding to each panel.  Note how the model is using the entire spectrum to derive abundance information, which includes both strong lines and the continuum, as opposed to just a subset of known strong lines.  See next page for remaining eight labels.} 
    \label{fig:coeffs}
\end{figure*}

\appendix
We present the first order coefficients of our \textit{Cannon} model as a function of wavelength in Figure A\ref{fig:coeffs}.  This figure illustrates the importance of each pixel when inferring various stellar parameters and abundances.  Each panel corresponds to a different label, and dashed red and gray lines mark strong lines of the element (where applicable) and Balmer lines.  While \textit{The Cannon} uses the entire spectrum to infer labels, strong lines of each element tend to have the largest coefficient values, suggesting that our model draws label information primarily from the expected strong lines.

Table A\ref{tab:fin} presents the table schema for our final catalog which contains \textit{Cannon}-rederived, higher precision parameters and abundances for 28,120 giant stars sampled by GALAH DR3.  We emphasize that these abundances are not necessarily accurate in the absolute sense.  Instead, they are accurate in the \textit{relative} sense: stars with the same spectra have the same labels.  We do not apply any global abundance offsets, as is common in surveys \citep[e.g.,][]{apogee, GALAHDR3}, to shift the scales of abundances.  In a practical sense, this catalog is useful for relative abundance comparisons, such as in studies of chemical homogeneity, but not for absolute comparison to theoretical yield models unless the comparison is relative in nature.  When using this catalog for most typical applications, we highly recommend only considering abundances with \code{chisq} and \code{X\_fe\_cannon\_chisq} < 2.  However, stars with \code{chisq} > 2 and abundances with \code{X\_fe\_cannon\_chisq} > 2 may be useful for identifying stars with unusual spectral features or anomalous abundances.

\begin{figure*}
    \centering
    \includegraphics[angle=90,origin=c, width=13cm]{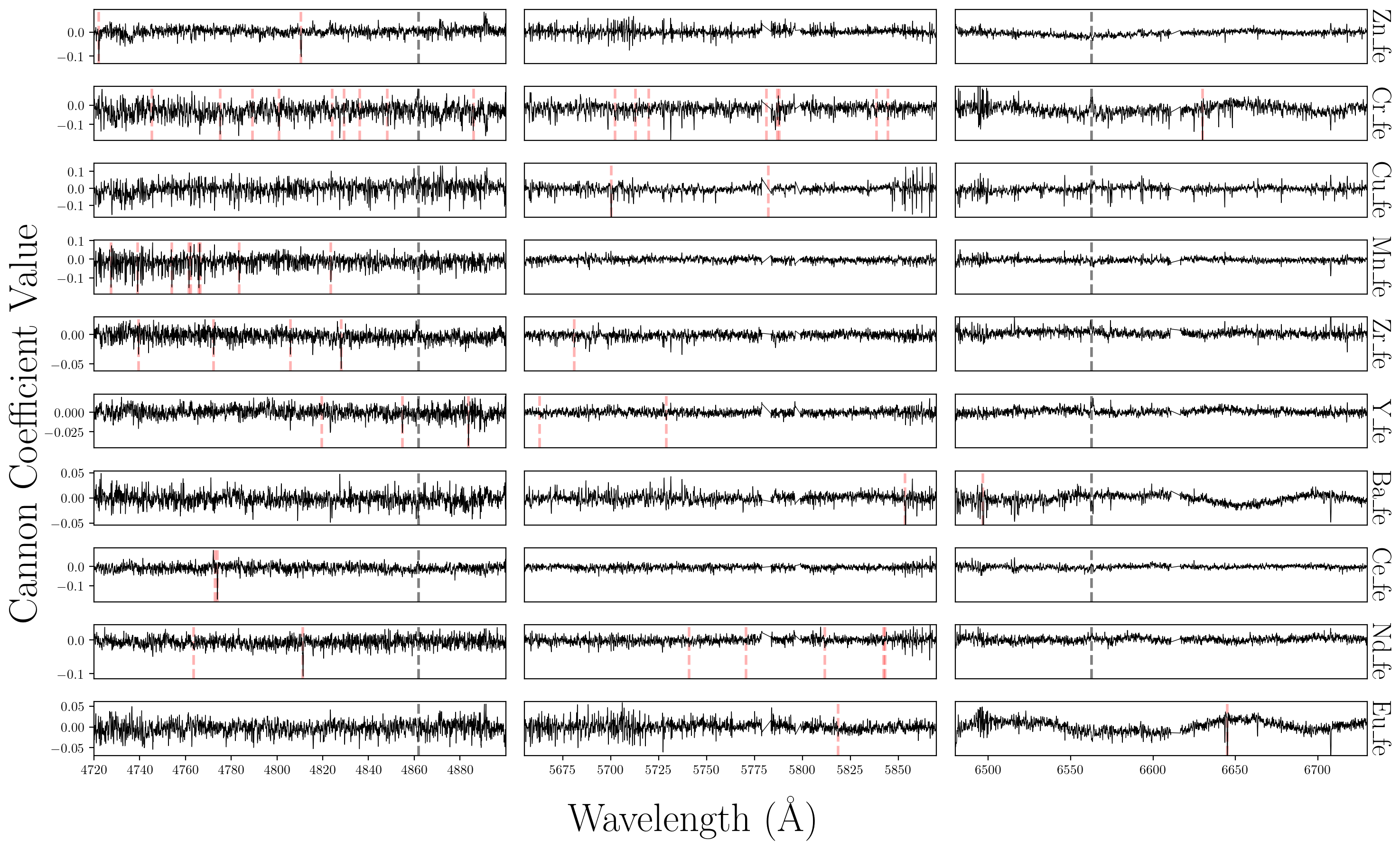}
    \caption{First order \textit{Cannon} model coefficients for the remaining eight labels.  See previous page for more details.}
    \label{fig:enter-label}
\end{figure*}

\begin{table*}
\begin{tabular}{cccl}
\hline
Column & Type & Unit & Description \\
\hline
\hline
\code{GaiaDR3\_source\_id} & \code{int} &  & Gaia DR3 source identifier (adopted from GALAH DR3) \\ 
\code{sobject\_id} & \code{int} &  & Unique GALAH identifier (adopted from GALAH DR3) \\ 
\code{chisq} & \code{float} &   & Global $\chi^2$ fit of \textit{The Cannon} model to GALAH spectrum \\ 
\code{vbroad\_cannon} & \code{float} & km s$^{-1}$  & \textit{The Cannon}-reported broadening velocity \\ 
\code{teff\_cannon} & \code{float} & K & \textit{The Cannon}-inferred $\rm T_{eff}$ \\ 
\code{logg\_cannon} & \code{float} &  & \textit{The Cannon}-inferred log g \\ 
\code{fe\_h\_cannon} & \code{float} &   & \textit{The Cannon}-inferred [Fe/H] \\ 
\code{e\_fe\_h\_cannon} & \code{float} &   & Final \textit{The Cannon} [Fe/H] uncertainty \\ 
\code{X\_fe\_cannon} & \code{float} &  & \textit{The Cannon}-inferred [X/Fe] (where X = O, Al, Mg, Ca, Si, Zn, Cu, Cr, \\ 
& & & Mn, Zr, Y, Ba, Ce, Nd, or Eu) \\
\code{e\_X\_fe\_cannon} & \code{float} &  & Final \textit{The Cannon} [X/Fe] uncertainty (where X = O, Al, Mg, Ca, Si, \\ 
& & & Zn, Cu, Cr, Mn, Zr, Y, Ba, Ce, Nd, or Eu) \\
\code{X\_fe\_cannon\_chisq} & \code{float} &  & Element-specific median $\chi^2$ fit of \textit{The Cannon} model to GALAH spectrum\\
& & & around strong lines of element \\
\hline

\end{tabular} \caption{Table schema for our full table containing our \textit{Cannon}-inferred parameters and abundances for the 28,120 stars in our sample.  Full table can be found at \code{zenodo.org}.}\label{tab:fin}
\end{table*}



\end{document}